\documentclass[twocolumn,aps,prc,superscriptaddress,floatfix,nofootinbib]{revtex4-2}
\usepackage[T1]{fontenc}
\usepackage{dsfont}               
\usepackage{mathrsfs}             
\usepackage{units}
\usepackage{slashed}              
\usepackage{amsmath}
\usepackage{amssymb}
\usepackage{amsbsy}
\usepackage{amsfonts}
\usepackage{epsfig,graphics}
\usepackage{graphicx}
\usepackage{dcolumn,multirow}
\usepackage{bm}
\usepackage[
colorlinks=true,
linkcolor=blue,
breaklinks=true,
urlcolor=blue,
citecolor=blue]{hyperref}
\usepackage{amsmath}
\usepackage[usenames]{color}
\usepackage{ulem} 
\usepackage{comment}
\usepackage{lineno}
\usepackage[percent]{overpic}
\RequirePackage[colorinlistoftodos,prependcaption,textsize=scriptsize,backgroundcolor=orange!20]{todonotes}
\newcommand{\er}{$\pm$}
\newcommand{\bc}           {\begin{center}}
\newcommand{\ec}           {\end{center}}
\newcommand{\bq}           {\begin{eqnarray}}
\newcommand{\eq}           {\end{eqnarray}}
\newcommand{\be}           {\begin{eqnarray}}
\newcommand{\ee}           {\end{eqnarray}}
\newcommand{\bi}           {\begin{itemize}}
\newcommand{\ei}           {\end{itemize}}

\newcommand{\oo}{$^o$}
\begin{document}

\title{\boldmath Decays of $N^*$ and $\Delta^*$ resonances into $N\rho$, $\Delta\pi$, and $Nf_0(500)$}

\newcommand*{\HISKP}{Helmholtz--Institut f\"ur Strahlen--
                     und Kernphysik, Universit\"at Bonn, 53115 Bonn, Germany}
\newcommand*{\JLAB}{Thomas Jefferson National Accelerator Facility, Newport News, Virginia 23606, USA}
\newcommand*{\FSU}{Florida State University, Tallahassee, Florida 32306, USA}

\affiliation{\HISKP}
\affiliation{\JLAB}
\affiliation{\FSU}

\author{A.V.~Sarantsev} \affiliation{\HISKP}
\author{E.~Klempt} \thanks{Corresponding author: \texttt{klempt@hiskp.uni-bonn.de}} \affiliation{\HISKP}\affiliation{\JLAB}
\altaffiliation{Corresponding author: \texttt{klempt@hiskp.uni-bonn.de}}
\author{K.V.~Nikonov} 
\affiliation{\HISKP} 
\author{T.~Seifen} 
\affiliation{\HISKP} 
\author{U.~Thoma} \affiliation{\HISKP} 
\author{Y.~Wunderlich}\affiliation{\HISKP}
\author{P. Achenbach}\affiliation{\JLAB}
\author {V.D.~Burkert} \affiliation{\JLAB}
\author {V.~Mokeev}\affiliation{\JLAB}
\author {V.~Crede} \affiliation{\FSU}

\date{\today}

\begin{abstract}
The decays of $N^*$ and $\Delta^*$ resonances into $N\rho$, $\Delta\pi$, and $Nf_0(500)$ final states are 
investigated in a coupled-channel analysis of data from pion- and photo-induced reactions. 
Improvements in the fit quality are observed upon the inclusion of additional resonance contributions. 
Branching ratios for the intermediate isobars $\Delta(1232)\pi$, $N\rho$, and $Nf_0(500)$ are reported.
\end{abstract}

\maketitle
\section{Introduction}
Photoproduction of mesons off protons or neutrons has become an important tool in the search for new 
$N^*$ and $\Delta^*$ resonances and for the study of their properties. A decisive factor in this progress has been the availability 
of high-precision data on a large variety of final states, produced with unpolarized or polarized photon beams incident 
on targets that may also be polarized. Several new excited states have been discovered, and many new properties of the 
$N^*$ and $\Delta^*$ resonances have been reported~\cite{Anisovich:2011fc,CBELSATAPS:2014wvh,CBELSATAPS:2015kka,Hunt:2018wqz,CBELSATAPS:2019ylw,CBELSATAPS:2020cwk,Ronchen:2022hqk}, and are now included in recent editions of the Review of Particle Physics (RPP)\cite{ParticleDataGroup:2024cfk}. Recent reviews on light-quark baryons can be found in Refs.~\cite{Klempt:2009pi,Crede:2013kia,Thiel:2022xtb,Burkert:2022adb,Gross:2022hyw}.

In this paper, we present results from the first coupled-channel analysis that incorporates data on pion- and photo-induced 
production of two neutral~\cite{CBELSATAPS:2015kka,CBELSATAPS:2022uad} and two charged pions~\cite{CLAS:2018drk,CLAS:2024iir,Crede:2024tbd}, 
using an event-based likelihood fit. New resonances are introduced; their impact on the fit will be reported elsewhere. Here, we present 
the branching ratios (BR's) of $N^*$ and $\Delta^*$ resonances decaying into $N\rho$, $\Delta\pi$, and $Nf_0(500)$.

The paper is organized as follows. First, we give a survey on earlier results on two-pion production
(Sec.~\ref{SectionOldData}). In Sec.~\ref{SectionNewData}, we present a survey of the data
used in our coupled-channel analysis. An outline of the coupled-channel formalism is given
in Sec.~\ref{SectionPWA}.  The results on masses, widths, helicity amplitudes, and BR's for decay
modes of $N^*$ and $\Delta^*$ resonances into $N \rho$, $\Delta\pi$ and $Nf_0(500)$ 
are presented in Sec.~\ref{SectionResults}. The paper ends with a
summary (Sec.~\ref{SectionSummary}).

\section{Previous results on two-pion production}
\label{SectionOldData}
\subsection{\boldmath Data on $\pi^- {p}\to {N}\pi\pi$}

Early studies of the dynamics of the $N\pi\pi$ final state were based on bubble
chamber data on the reaction $\pi^- {p}\to {N}\pi\pi$ from the
Lawrence Berkeley National Laboratory, the Laboratoire National Henri Becquerel at Saclay, and 
the  Rutherford Appleton Laboratory. Several groups
reported results from their partial-wave analyses. Here, we restrict the discussion to those analyses of bubble chamber
data~\cite{Manley:1984jz,Longacre:1974xu,Longacre:1977ga,Barnham:1980za,Vrana:1999nt} which were
included in the Review of Particle Properties~\cite{ParticleDataGroup:2014cgo}.
Manley, Arndt, Goradia, and Teplitz \cite{Manley:1984jz} collected the existing data on $\pi^-p \to
n\pi^+\pi^-$, $\pi^- p\to p\pi^0\pi^-$, $\pi^+p \to n\pi^+\pi^+$ and $\pi^+p \to p\pi^+\pi^0$.
References to the data can be found in their paper. The 241,214 events were grouped into 22 bins
in energy in the mass range from 1320 to 1930\,MeV. The width of the bins varied from 20 to 40\,MeV
depending on the available statistics.  In an event-based likelihood fit, the partial wave amplitudes
for $\pi N\to N\rho $, $\pi N \to Nf_0(500) $, $\pi N \to \Delta(1232)\pi$, and $\pi N \to N(1440)\nicefrac{1}{2}^+\pi$ were constructed
in slices of the invariant mass. Note that for resonances with a total spin $J>\nicefrac12$, there are two $\Delta\pi$- amplitudes, and
three $\pi N\to N\rho $ partial-wave amplitudes with two different orbital angular momenta $L$ between $\rho$ and $N$. In one of the cases the 
$\rho$-spin $S$ is aligned parallel, in the other one antiparallel to the nucleon spin. 
Considering waves up to angular momenta $L\leq4$, 76 $\Delta\pi$-
$N\rho$- and $Nf_0(500)$-waves can thus be constructed. After preliminary fits, 32 of
them were retained for the final analysis.  The partial-wave amplitudes are
presented as partial-wave cross sections and as Argand diagrams. Energy-dependent fits identified nine $N^*$
and five $\Delta^*$ resonances and their decays into $\Delta(1232)\pi$ and $N\rho$. The decays of
nine $N^*$ resonances into $Nf_0(500)$ were reported. 

Similar analyses had been carried out by Longacre {\it et al.} \cite{Longacre:1974xu,Longacre:1977ga}. These
analyses were based on a smaller statistical sample, and a smaller number of partial-wave amplitudes were
extracted from the data. No uncertainties were given for the final results. Given the intrinsic uncertainties of
the complex analysis, the consistency of the results is fair even though some significant differences are seen
between the results from Refs.~\cite{Longacre:1974xu,Longacre:1977ga}. 
A large number of resonances were found to contribute to
the reaction $\pi N\to N\pi\pi $.  The reaction $\pi^+p\to N\pi\pi$ was studied in~\cite{Barnham:1980za} for resonances
with masses below 1700\,MeV. The main result was the first observation of $\Delta(1600)\nicefrac32^+\to \pi N(1440)\nicefrac12^+$
decays.

Vrana, Dytman, and Lee~\cite{Vrana:1999nt} combined the partial-wave amplitudes for the transitions from $\pi N$
to eight different final meson-baryon states in a coupled channel formalism. For the $\pi N\to N\pi\pi $ transitions, the
authors exploit the quasi-two-body channel decomposition of Manley, Arndt, Goradia, and Teplitz~\cite{Manley:1984jz}.
The results on $N^*$ and  $\Delta^*\to N\pi\pi$ decays are mostly 
compatible with the results reported in~\cite{Manley:1984jz}.

New experimental input came when the Crystal Ball experiment was moved to Brookhaven National Laboratory
(BNL). This extremely successful and enduring detector was built at the Stanford Linear Accelerator to
detect $\gamma$-ray transitions in the charmonium system~\cite{Bloom:1983pc}. Subsequently it was moved to
DESY for studies of two-photon collisions and $\Upsilon$ spectroscopy \cite{Bienlein:1991jb}. Then it
returned to BNL for baryon spectroscopy~\cite{Nefkens:2005an}, and continued to work in this field
when it was installed at the Mainz Microtron (MAMI) \cite{Denig:2016dqo}. 

At BNL, precision data on $\pi^-p\to n\pi^0\pi^0$  were measured covering the mass region of
the Roper $N(1440)\nicefrac12^+$ resonance~\cite{CrystalBall:2004qln}.  The data led to a series of papers 
that tried to improve our understanding of $N(1440)\nicefrac12^+$ 
\cite{Kamano:2006vm,Schneider:2006bd,Siemens:2014pma,Shklyar:2014kra}. The study of~\cite{Kamano:2006vm} was extended to higher masses
in~\cite{Kamano:2008gr}; later, the total cross section was decomposed into partial-wave
contributions \cite{Kamano:2013ona}.
These analyses used mass projection and angular distributions for the fits, a qualitative description
of the data was achieved.

Recently, the second resonance region was studied by the HADES collaboration~\cite{HADES:2020kce}
and $\Delta\pi$, $Nf_0(500)$ and $N\rho$ excitation functions were determined.

\subsection{\boldmath Data on $\gamma {p}\to {N}\pi\pi$}
Multipion photoproduction off protons was studied with bubble chambers at Cambridge
\cite{CambridgeBubbleChamberGroup:1968zz}, DESY \cite{Aachen-Berlin-Bonn-Hamburg-Heidelberg-Munchen:1968dpq}, and at SLAC \cite{Ballam:1971wq,Ballam:1971yd}.
The multiplicities of charged pions, cross sections and $N\pi$, $\pi\pi$ and three-pion invariant 
mass distributions were determined for incident
photons of up to $E_\gamma =8$\,GeV. In Ref.~\cite{Lanzerotti:1965bp,Lanzerotti:1968zz}, a detailed 
study of the shape of the $\rho$-meson was reported.
The maximum of the $\pi^+\pi^-$ effective mass
distribution was observed to be shifted with respect to the nominal $\rho$ mass;
the asymmetric shape was interpreted by S\"oding~\cite{Soding:1965nh} as interference of the amplitude for
diffractive  $\rho$-meson production with the amplitude in which the photon dissociates into 
$\pi^+$ and $\pi^-$, and one of the pions is elastically scattered off the proton. A contact (Kroll-Rudermann) term
 \cite{Kroll:1954} dominates the leading $\Delta^{++}\pi^-$ production at low energies.

A comprehensive study of various final states in photoproduction was performed at DESY in a streamer
chamber experiment~\cite{Aachen-Hamburg-Heidelberg-Munich:1975jed}. The energy of the photons (1.6\,GeV $< E_\gamma <$
6.3\,GeV) was tagged by a momentum measurement of the deflected high-energy positron. Therefore, 
final states with one neutral particle could be reconstructed. 

In 1974, the $J/\psi$ was discovered \cite{E598:1974sol,SLAC-SP-017:1974ind}, and the interest of the
high-energy community turned to the charmonium spectrum and into $J/\psi$ decays \cite{Kopke:1988cs}.
On the other hand, the development of chiral perturbation theory by Weinberg~\cite{Weinberg:1978kz} and
Gasser and Leutwyler~\cite{Gasser:1983yg} opened new opportunities to understand QCD in its low-energy regime,
and the production and decays of baryon resonances became a topical theme of hadron physics.

A number of experiments on different charge modes of $\gamma p \to N \pi\pi$ was performed
at MAMI at photon energies from the two-pion production threshold to $E_\gamma=800$\,MeV
\cite{Braghieri:1994rf,Harter:1997jq,Wolf:2000qt,Langgartner:2001sg,Kotulla:2003cx,%
GDH:2005jgl,GDH:2007nkn,CrystalBallatMAMI:2009iym}. The total and differential cross sections,
$d\sigma/dM(p\pi)$, $d\sigma/dM(\pi\pi)$  and Dalitz plots, as well as beam-helicity asymmetries,
were measured for the three isobar channels. 
The data provided for an improved knowledge
of two-pion decay modes of $N(1440)\nicefrac12^+$ and $N(1520)\nicefrac12^+$ \cite{GomezTejedor:1995pe,GomezTejedor:1995kj}.

Later, MAMI was upgraded with
a further acceleration stage to a maximal electron energy of 1604\,MeV, and the reactions
$\gamma p \to p\pi^0 \pi^0$ and $\gamma n \to n\pi^0 \pi^0$ were studied
\cite{Kashevarov:2012wy,Zehr:2012tj,Oberle:2013kvb,A2:2014snn,A2:2015pgk}. Cross sections
and invariant-mass distributions as well as beam helicity asymmetries were reported.

The GRAAL collaboration measured the total cross section, the differential cross
sections $d\sigma/dM(p\pi^0)$, $d\sigma/$ $dM(\pi^0\pi^0)$, and the
beam asymmetry $\Sigma$ for $\gamma p\to p\pi^0\pi^0$ in the beam energy
range $0.65<E_\gamma< 1.5$\,GeV~\cite{Assafiri:2003mv}. A similar study
was conducted by the SAPHIR Collaboration for $\gamma p\to p\pi^+\pi^-$ \cite{Wu:2005wf}.
The total and differential cross sections were determined for photon energies up to 2.6\,GeV,
and the contributions from $p\rho^0$, $\Delta(1232)^{0}\pi^+$ and $\Delta(1232)^{++}\pi^-$
were identified. At high photon energies, the $s$-channel
helicity is conserved, but the data suggest that at lower energies, $t$- or
$u$-channel exchanges also contribute to the reaction.

The Crystal Barrel detector \cite{CrystalBarrel:1992qav} was built to study mesons produced in $\bar pp$
annihilation at the CERN Low Energy Antiproton Ring LEAR \cite{Klempt:2002ap,Klempt:2005pp,Bugg:2004xu,Amsler:2019ytk}.
After completion of the experimental program at LEAR, the detector was moved to Bonn and in 2000 
installed at ELSA.

The first studies of the reaction $\gamma p\to p\pi^0\pi^0$ covered the photon energy range
from 0.4 to 1.3\,GeV. The reaction was analyzed jointly with the $\pi N$ scattering
amplitudes~\cite{Arndt:2006bf} derived from $\pi N$
$\pi^\pm p$ elastic, pion charge exchange, and $\pi^- p
\to \eta n$. Furthermore, a large set of data on pion- and photo-induced reactions available
at that time was included. In an event-based partial wave
analysis, properties of the Roper resonance were deduced: mass, width, helicity coupling,
and partial decay widths for decays into $N\pi$, $\Delta(1232)\pi$, and
$pf_0(500)$~\cite{Sarantsev:2007aa}. In addition, the branching ratios of the
resonances in the second and third resonance regions for decays into
$\Delta(1232)\pi$, $pf_0(500)$, $N(1440)\pi$, and $N(1520)\pi$ were
determined \cite{Thoma:2007bm}. Later, the energy range was extended to cover the
photon energies from 600\,MeV to 2.5\,GeV.  The most prominent are the decays that occur
through $\Delta(1232)\pi$, $N(1440)\pi$, $N(1520)\pi$, and $N(1680)\pi$, but also $pf_0(500)$ and
$pf_0(980)$ were observed \cite{CBELSATAPS:2015kka}. Two aspects were particularly
important: a) The distributions of the polarization observables $I^s$ and $I^c$, characterizing correlations between a linear photon polarization and the direction of outgoing single particles in photoproduction of three-body final states, suggest that, in the 1.8 to 2.0 GeV mass region, the $N(1520)\nicefrac32^-\pi$ intermediate state is reached from a
dominant $J^P=\nicefrac32^+$ wave. The resonance $N(1900)\nicefrac32^+$ is thus confirmed without partial wave analysis, in a
completely model-independent way~\cite{CBELSATAPS:2015tyg}. b) It is shown that $\Delta(1910)\nicefrac12^+$, $\Delta(1920)\nicefrac32^+$, $\Delta(1905)\nicefrac52^+$, $\Delta(1950)\nicefrac72^+$ and the corresponding
spin-parity series in the nucleon sector, $N(1880)\nicefrac12^+$, $N(1900)\nicefrac32^+$, $N(2000)\nicefrac52^+$, and $N(1990)\nicefrac72^+$ all
decay into $\Delta(1232)\pi$, but only the nucleon resonances decay significantly via $N(1440)\nicefrac12^+\pi$,
$N(1520)\nicefrac32^-\pi$, $N(1680)\nicefrac52^+\pi$, and $Nf_0(500)$ \cite{CBELSATAPS:2015kka}. The latter decay modes 
proceed via isobars having intrinsic orbital excitations. The four $\Delta^*$ resonances are compatible with a 3-quark-wave function where always only one of the two oscillators is excited at a certain time, the four $N^*$ resonances
have a component in their wave function in which two oscillators are excited at the same time. Apparently, this component
first de-excites one of the two oscillators into an isobar where one oscillator is still excited. The results
thus provide evidence that the spectrum of $N^*$ resonances takes full advantage of the dynamics of a three-body
system~\cite{CBELSATAPS:2015taz}. A study with increased statistics confirmed these findings~\cite{CBELSATAPS:2022uad}.

\begin{figure}[pb]
\centering
    \includegraphics[width=0.99\linewidth, trim= 0mm 3mm 0mm 0mm, clip]{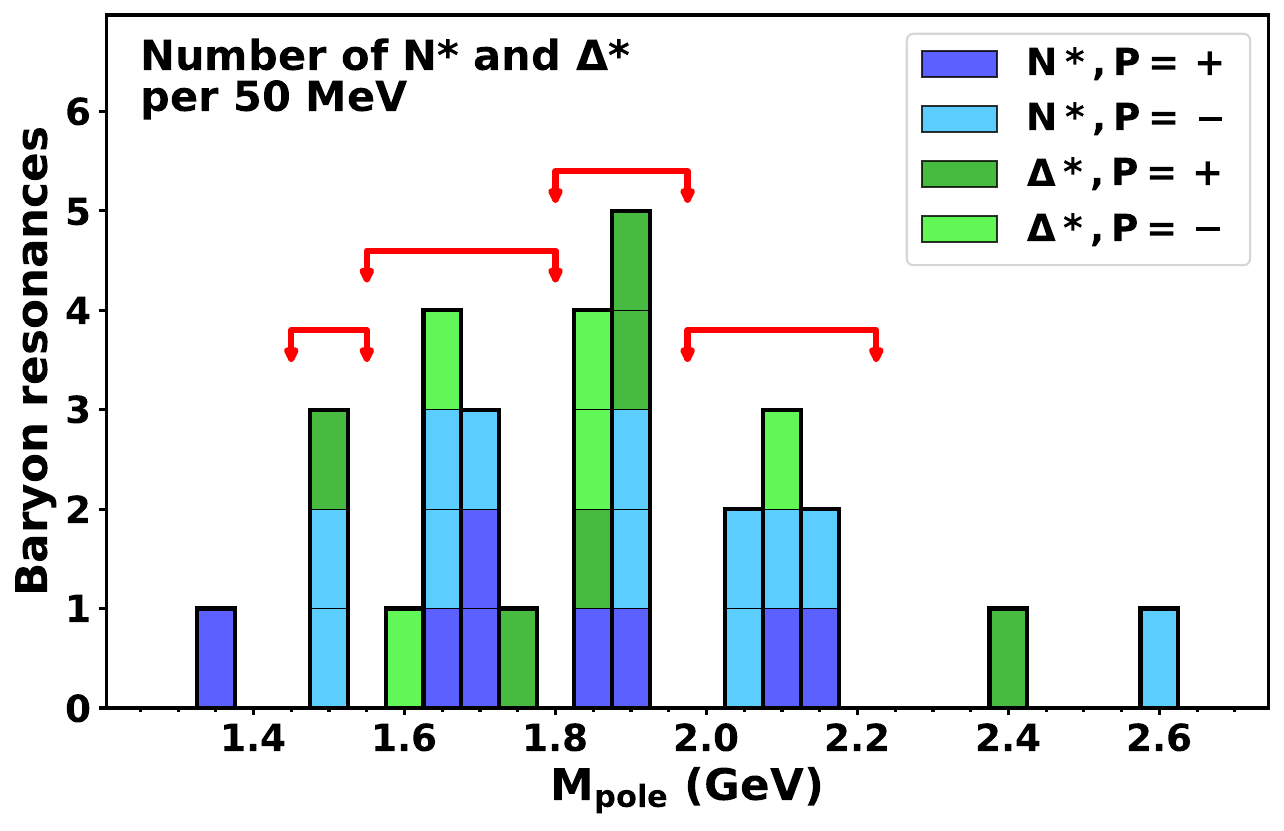}
\caption{\label{fig:shells}The number of 3* and 4* $N^*$ and $\Delta^*$ resonances in~\cite{ParticleDataGroup:2024cfk} above the ground states in 50\,MeV bins.
The arrows mark the mass range for the Dalitz plots shown below. The second, third, fourth, and fifth resonance
regions are covered.}
\end{figure}

The CLAS Collaboration reported measurements of the beam-helicity asymmetry
$I^\odot$ for $\gamma p\to p\pi^+\pi^-$~\cite{CLAS:2005oqk} for center-of-mass energies between
1.35 and 2.3\,GeV. The dynamics of the reaction $\gamma p\to p\pi^+\pi^-$
was studied in the mass range from 1.6 to 2.0\,GeV in Ref.~\cite{CLAS:2018drk}, and
results on resonance photocouplings were determined.
The analysis was extended up to 2.6\,GeV in~\cite{CLAS:2024iir},
and excitation curves and density matrix elements were reported.
Additional new results exploiting the polarization
variables included in this analysis are
presented in~\cite{Crede:2024tbd}.

\begin{figure*}
     \centering\vspace{-5mm}
\includegraphics[width=\linewidth]{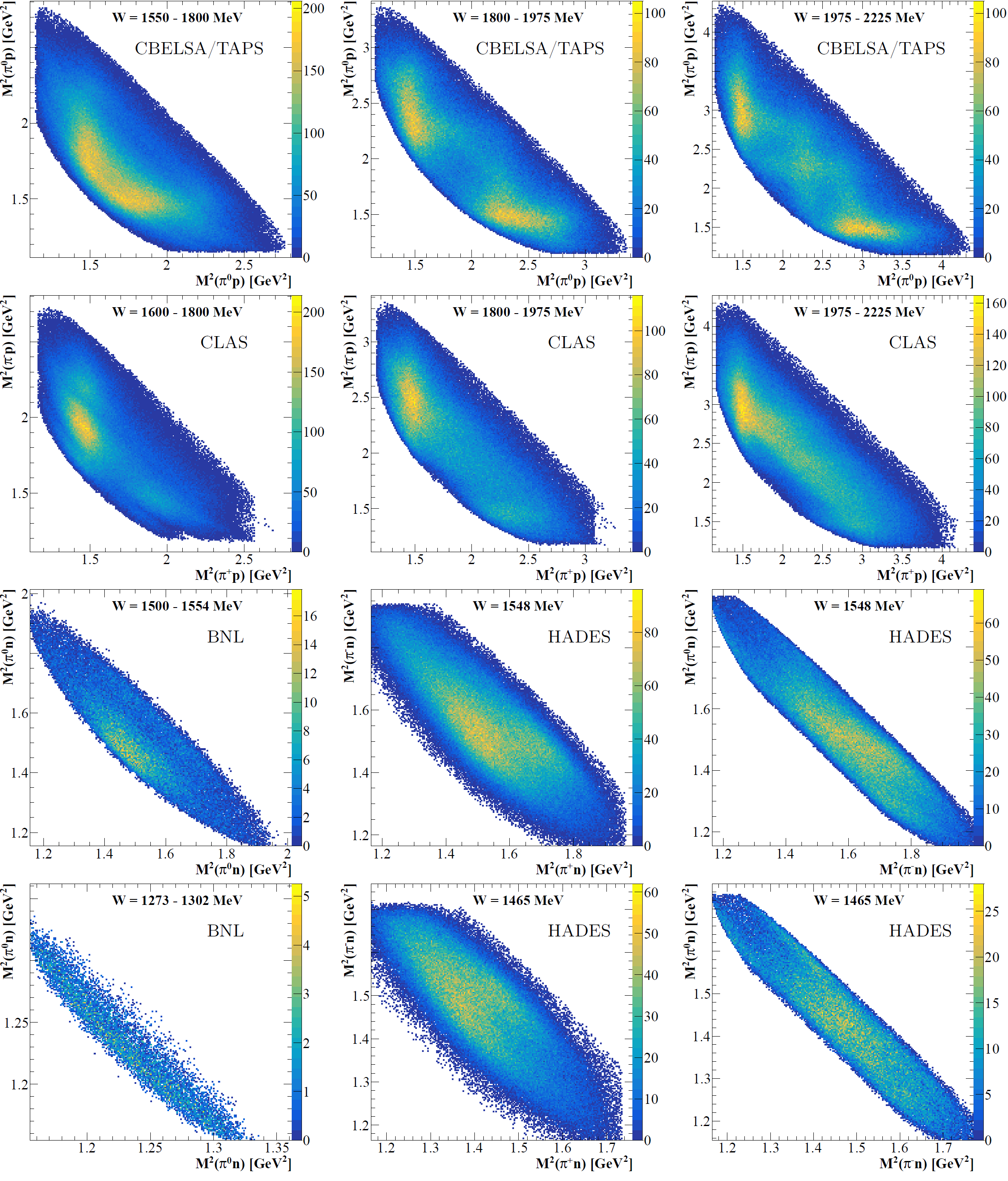}
     \caption{(top) CBELSA/TAPS data \cite{CBELSATAPS:2015kka} on $\gamma p\to \pi^0\pi^0 p$ for different ranges of $\sqrt s$.
     (second row) CLAS data \cite{CLAS:2024iir} on $\gamma p\to \pi^+\pi^- p$. (third and fourth row, left) 
     BNL data \cite{CrystalBall:2004qln} on $\pi^- p\to \pi^0\pi^0 n$, (middle) HADES data \cite{HADES:2020kce} on $\pi^- p\to \pi^+\pi^- n$,
     (right) HADES data \cite{HADES:2020kce} on $\pi^- n\to \pi^0\pi^- n$.  BNL has reported data for six mass ranges,
      HADES for four ranges, the lowest and highest mass ranges are shown.  The data are shown without acceptance correction and flux normalization.  \vspace{-4mm}}
         \label{fig:DP_CBELSA/TAPS}
\end{figure*}

This paper is devoted to the study of cascade decays of $N^*$ and $\Delta^*$ resonances decaying
into $\Delta\pi$, $N\rho$, and $Nf_0(500)$. 
Figure~\ref{fig:shells} shows the number of states in the second,
third, fourth, and fifth resonance regions. 
The (red) range markers indicate the ranges for
which Dalitz plots of $N\pi\pi$ final states for data from the CBELSA/TAPS~\cite{CBELSATAPS:2015kka} and the CLAS~\cite{CLAS:2024iir} experiment, the Crystal Ball experiment at BNL~\cite{CrystalBall:2004qln}, and   HADES~\cite{HADES:2020kce} are shown in Fig.~\ref{fig:DP_CBELSA/TAPS}.

\section{Data used in the coupled-channel analysis}
\label{SectionNewData}

\subsection{The database}
The Bonn-Gatchina partial wave analysis is based on a large data base. Nearly all available data on
pion- and photo-induced reactions are taken into account in a coupled channel analysis.
The main data sets include photo-induced reactions of protons and neutrons into
$N\pi$, $N\eta$, $N\eta'$, $K\Lambda$, $K\Sigma$, $N\pi\pi$, and $N\pi^0\eta$, including
polarization data when available. The analysis is constrained by the real and imaginary
parts of $\pi N$ elastic scattering amplitudes either from the Karlsruhe analysis KA84~\cite{Koch:1985bp}
or from the GWU analysis WI08~\cite{Workman:2012hx}. 
$\pi^\pm$ inelastic scattering off protons into $K\Lambda$ and $K\Sigma$ final states including
data with a polarized target are analyzed as well.  
A complete list of data
used in this analysis can be found on our web site~\cite{BnGa-web}. The solution presented here is named BnGa-2024.

\subsection{Data on two-pion production}
The main emphasis of this paper are data on $\gamma p\to \pi^+\pi^- p$ {with unpolarized photons and protons} \cite{CLAS:2018drk,CLAS:2024iir}
and with linear photon and transverse target polarization~\cite{Crede:2024tbd}, data on
$\gamma p\to p\pi^0\pi^0$ without polarization~\cite{Zehr:2012tj,Kashevarov:2012wy,CBELSATAPS:2015kka},
with linearly polarized photons without target polarization \cite{CBELSATAPS:2015tyg}, and with linearly polarized photons and 
transversely polarized target protons~\cite{CBELSATAPS:2022uad}.
Also included in the data base are data on the pion-induced reactions $\pi^-p\to n\pi^0\pi^0$~\cite{CrystalBall:2004qln} from BNL,
$\pi^-p\to n\pi^+\pi^-$~\cite{HADES:2020kce}, and $\pi^-p\to p\pi^-\pi^0$~\cite{HADES:2020kce} from GSI, and $\gamma p\to p\pi^0\pi^0$ 
with linearly polarized photons from GRAAL~\cite{Ajaka:2007zz}.

Some of the data are shown in the form of Dalitz plots, see Figs.~\ref{fig:DP_CBELSA/TAPS}. Note that the CLAS data 
consist of 400 million events \cite{CLAS:2018drk} available
in the form of nine mass and angular distributions; only 1.842 million events are available event-wise and can be shown
as Dalitz plots. Both data sets are used in the fits described below.

\section{Partial-wave analysis method}
\label{SectionPWA}
\subsection{The amplitudes}
In the BnGa approach, the energy-dependent part of
the photo-production amplitude is described by a $D$-matrix which is
based on dispersion relations. The amplitude is
covariant and preserves analyticity and two-body unitarity. The
block $D_{\alpha\beta}$, describing a transition between the
bare state $\alpha$ and the bare state $\beta$, is calculated by
solving the equation
\be
D_{\alpha\beta}= D_{\alpha\gamma}\sum\limits_j
B^j_{\gamma\eta}d_{\eta} \delta_{\eta\beta}+d_{\alpha}\delta_{\alpha\beta}.
\label{D-comp}\vspace{-1mm}
\ee
Here, $\delta$ is the Kronecker symbol, the amplitudes $d$  
represent resonant and non-resonant terms, and
$B^j_{\gamma\eta}$ meson-baryon loop diagrams.
The summation $j$ takes into account all relevant  intermediate states. 
For example,
in the $K$-matrix approach, the bare states representing resonances correspond to $K$-matrix poles and non-resonant 
contributions. The non-resonant contributions include t- and u-exchange amplitudes. 
These non-resonant terms  can not have singularities in the physical region but only so called left-hand side singularities, located in the region of logarithmic singularities of the original amplitudes. 
In matrix form, Eq.~(\ref{D-comp}) can be written as\vspace{-1mm}
\be
\hat D= \hat D\hat B\hat d+\hat d, \quad \hat D= \hat d(\mathds{1}-\hat B\hat
d)^{-1},\quad \hat d_{\alpha\beta}=d_\alpha\delta_{\alpha\beta}.
 \ee
The $\hat d$ is a diagonal matrix which collects the bare-state propagators (poles) as well as terms describing the non-resonant transitions:\vspace{-1mm}
 \be
 \hat d=\text{diag}\left
(\frac{1}{M^2_1-s},\ldots,\frac{1}{M^2_n-s},R_1,R_2 \ldots\right),\vspace{-1mm}
\ee
where $n$ is the number of poles and $R_\alpha$ are propagators for non-resonant transitions (for
example, for left-hand poles or constants). The elements of the $\hat B$-matrix
correspond to the sum of meson-baryon loop diagrams\vspace{-1mm}
\begin{eqnarray}
\hat B_{\alpha\beta}&=&\sum\limits_j
B^j_{\alpha\beta} \nonumber \vspace{-1mm}\\
B^j_{\alpha\beta}&=&\int\frac{ds'}{\pi}\frac{g^{L(\alpha)}_j\varrho(s',m_{1j},m_{2j})W_j(s)g^{R(\beta)}_j}{s'-s-i0}
\;.\label{bf}\vspace{-1mm}
\end{eqnarray}
The $g^{R(\beta)}_j$ and $g^{L(\alpha)}_j$ are right and left
vertices for a transition from the state $\alpha$ to state $\beta$ via the intermediate 
meson-baryon loop $j$.  The $\varrho(s,m_{1j},m_{2j})$ represents the standard phase volume,
normalized to 1 at $s\to\infty$:\vspace{-1mm}
\be
\varrho(s,m_{1},m_{2})=\frac{\sqrt{(s\!-\!(m_{1}\!+\!m_{2})^2)(s\!-\!(m_{1}\!-\!m_{2})^2)}}{s}\,\,
\ee
This phase volume corresponds to the imaginary part of the loop diagram for scattering two spinless particles in S-wave. The factor $W_j(s)$ describes an additional energy factor which comes from spin-orbital structure of the decay vertex for the partial wave considered and the propagator structures of the final particles. Usually this factor is proportional to $|\vec k|^{2L}$ where $L$
is the orbital momentum of the decay channel and $\vec k$ is the relative momentum
of the final particles calculated in the c.m.s. of the reaction.
To compensate this divergence at high energies, we include
Blatt-Weisskopf form-factors with radius 0.8\,fm into the definition of the $W_j (s)$.
The exact expression for the factors $W_j$ for different two-particle and
three particle channels can be found in \cite{Anisovich:2006bc}.
Even after including the Blatt-Weisskopf form-factors, the integral eq.(\ref{bf})  diverges,
and we regularize it by a one-step subtraction procedure.

For the pole terms the left and right-side couplings
coincide:\vspace{-1mm}
\be
g^{R(\beta)}_j=g^{L(\alpha)}_j=g^{(\alpha)}_j\,,\ j=1,\ldots,m\vspace{-1mm}
\label{ver_pole}
\ee
where $m$ is the number of the decay channels. The non-resonant
terms have a more complicated parametrization. For example, if we
introduce a non-resonant transition from $\pi N$ to all
other channels we need to add two bare terms with the vertices:\vspace{-1mm}
\begin{eqnarray}
g^{L(n+1)}_{j=1,\ldots,m}&=&(f_{11},f_{12},\ldots,f_{1m})\nonumber \\
g^{R(n+1)}_{j=1,\ldots,m}&=&g^{L(n+2)}_{j=1,\ldots,m}=(1,0,\ldots,0)\nonumber \\
g^{R(n+2)}_{j=1,\ldots,m}&=&(0,f_{21},\ldots,f_{m1})
\label{eq7}
\end{eqnarray}
The transition amplitude $A_{ab}$ from the initial channel $a$ to the final channel $b$ is calculated by the convolution of the matrix $D_{\alpha\beta}$ with vectors formed from the production couplings $g_a^{R(\alpha)}$ and decay couplings $g_b^{L(\beta)}$ of the bare states: 
\be
A_{ab}=\sum\limits_{\alpha,\beta=1}^{n+2} g^{R(\alpha )}_a
D_{\alpha\beta}g^{L(\beta)}_b\,.
\ee
If all couplings are real numbers, these amplitudes obey
unitarity and analyticity. This equation is applied to calculate
transitions between different two-particle final states.

The transition amplitude from a two-body initial state into a
three-body final state has an additional
complexity due to rescattering of the final-state particles. In this case,
the decay into three-body channel '$b$' can be approximated as\\[-4ex]
\be
A_{ab}=\sum\limits_{\alpha,\beta=1}^{n+2}g^{R(\alpha )}_a
D_{\alpha\beta}G^{(\beta)}_b\,,
\vspace{-3mm}
\ee
where
\begin{eqnarray}
\label{decay}
&&G^{(\beta)}_b\!\!=\!g^{(\beta)}_b e^{i\phi^{(\beta)}_b} \quad {\rm for} \quad \beta=1,\ldots,n\,,\nonumber \\ && G^{(n+1)}_b\!=f_{1b}e^{i\phi_{1b}},\, \quad G^{(n+2)}_b\!\!=0.~~~
\end{eqnarray}
The phases $\phi^{(\beta)}_b$ and $\phi_{1b}$ are fit parameters.
The amplitude eq.(\ref{decay}) can also describe the transition from the initial state to those channels, which are not explicitly taken into account in the eq.(\ref{D-comp}). This is convenient when a channel contributes very little to the total cross section but may be important to describe 
the interference of amplitudes, e.g. in the fit to polarized data. In our parametrization, we explicitly include all fitted two-particle and 
three-particle channels (with different intermediate resonances) that contribute at least 10 MeV to the total width of any state 
in a given partial wave. 
In eq.(\ref{D-comp}) we also include an additional channel, the so-called ``missing inelasticity channel". It represents the 
inelasticity of all channels that are not taken into account explicitly. 
For resonances in the low- and intermediate-mass regions, 
the contribution of a missing inelasticity channel was found to be close to zero. However, for the states in the 1900\,MeV region and above,
it can describe up to about 50\% of the width of the observed states. The missing inelasticity channel is probably needed when resonances have
significant couplings to four (or more) particles like $\Delta\rho$.

The three-body phase volume can be calculated as a spectral integral
in the production of an intermediate resonance. For the $\rho
N$ intermediate state, the integral is given by
\be \varrho_{\rho N} (s)=
\!\!\!\!\!\!\!\int\limits_{4m_\pi^2}^{(\sqrt{s}-m_N)^2}
\!\!\!\!\frac{ds_{12}}{\pi} \frac{\varrho(s,\sqrt{s_{12}},m_N)W_{\rho N}
M_\rho\Gamma^\rho_{\pi\pi}}
{(M_\rho^2\!-\!s_{12})^2\!+\!(M_\rho\Gamma^{\rho}_{tot})^2}.~
\label{PSrhoN}
\ee
The explicit expression of the function $W_{\rho N}\equiv W_{\rho N}(s,s_{12},m_N)$
can be found in \cite{Anisovich:2006bc} as well as for other intermediate states. The remaining quantities appearing in equation \eqref{PSrhoN} are defined as:
\begin{eqnarray}
&&M_\rho\Gamma^\rho_{\pi\pi}=g^2_{\pi\pi}\frac{|\vec
q_\pi|^2}{3} \varrho(s_{12},m_\pi,m_\pi), \nonumber \\
&&|\vec q_\pi|^2=\frac{ \big
(s_{12}\!-\!4m_\pi^2\big )}{4}, \quad
\varrho(s_{12},m_\pi,m_\pi)=\frac{2|\vec q_\pi|}{\sqrt{s_{12}}}\;.~~~~~~~
\end{eqnarray}
Here, $g_{\pi\pi}$ is the coupling of the $\rho$-meson into two pion channel. The approach satisfies two-body unitarity and quasi-three body unitarity: The genuine three-body unitarity is compromised by excluding transitions between amplitudes with intermediate states in different kinematical channels: 
$N\rho\to \Delta(1232)\pi$ meson-baryon loops are, e.g., not included. 
\\[-2ex]

The photoproduction amplitude is 
described with the $P$-vector approach~\cite{Aitchison:1972ay,Chung:1995dx,Peters:2004qw}. The $\gamma N$ channel does not contribute to the loop 
diagrams, and the $\gamma N$ vertex is only taken into account once. In this case it is convenient to rewrite eq.(\ref{D-comp})  in the form
\be
\hat D=d(\mathds{1}-\hat B\hat d)^{-1}=
d+d(\mathds{1}-\hat B\hat d)^{-1}\hat B\hat d\,,
\ee
and the transition amplitude from the $\gamma N$ channel to the channel $b$ can be written as
\be
A_{b}\!=\!\sum\limits_{\alpha=1}^{n+2}P^{(\alpha)}d_{\alpha}G^{(\alpha)}_b\!+\!\!\!\sum\limits_{\alpha,\beta,\eta=1}^{n+2}
\!\!\!P^{(\alpha )}
D_{\alpha\beta}\hat B_{\beta\eta}d_{\eta} G^{(\eta)}_b\!.
\ee
The $P^{(\alpha)}$-vector ($\alpha\le n$) describes the $\gamma N$ couplings of the resonances. The non-resonant parts include
$t$- and $u$-channel amplitudes. The $u$-channel amplitudes are taken as amplitudes with 
the exchange of the corresponding baryon pole. The $t$-channel amplitudes are parameterized as reggeized pion-, $\rho (\omega)$, and Pomeron-exchange amplitudes. 
These are represented by the exchange of a Reggeon \cite{Sarantsev:2008ar} where the energy dependent part has the form
\bq
A&=&g(t)R(\xi,\nu,t) \quad {\rm where}\\
R(\xi,\nu,t)&=&\frac{1+\xi
\exp(-i\pi\alpha(t))}{\sin(\pi\alpha(t))} \left (\frac{\nu}{\nu_0}
\right )^{\alpha(t)} \;\nonumber
\eq
and the spin-orbital structure of vertices is given in \cite{Anisovich:2004zz}. 
We use $g(t)=g_0\exp(-bt)$ as the vertex function and form factor.
$\alpha(t)$  describes the trajectory, $\nu=\frac 12 (s-u)$, $\nu_0$
is a normalization factor, and $\xi$ the signature of the
trajectory. Pion and Pomeron exchange both have a positive
signature and therefore \cite{Anisovich:2004zz}:
\be
R(+,\nu,t)=\frac{e^{-i\frac{\pi}{2}\alpha(t)}} {\sin
(\frac{\pi}{2}\alpha(t))} \left (\frac{\nu}{\nu_0}\right
)^{\alpha(t)}\;.
\ee
To eliminate the poles at $t<0$, additional $\Gamma$-functions are
introduced in (\ref{eq9a}):
\be\label{eq9a}
\sin \left (\frac{\pi}{2}\alpha(t)\right ) \to \sin \left
(\frac{\pi}{2}\alpha(t)\right )  \; \Gamma \left (\frac
{\alpha(t)}{2}\right )\, .
\label{rho_1}
\ee
The $\rho(\omega)$-exchange amplitudes have a negative signature and were parameterized as:
\be
R(-,\nu,t)=\frac{e^{-i\frac{\pi}{2}\alpha(t)}} {\cos
(\frac{\pi}{2}\alpha(t))\Gamma \left (\frac
{\alpha(t)+1}{2}\right )} \left (\frac{\nu}{\nu_0}\right
)^{\alpha(t)}\!.
\ee
The pion, $\rho(\omega)$ and Pomeron trajectories were taken with the standard
parametrization:
\bq
\label{eq11a}
\hspace{-5mm}\pi\quad\qquad\qquad\alpha(t)&=-&0.25+0.85 ({\rm GeV}^{-2})t\\\label{eq11b}
\hspace{-5mm}\rho(\omega)\qquad\qquad\alpha(t)&=&0.50+0.85 ({\rm GeV}^{-2})t\\
\label{eq11c}
\hspace{-5mm}\text{Pomeron}\quad\quad\alpha(t)&=&0.26+0.85  ({\rm GeV}^{-2})t
\eq
where $t$ should be given in GeV$^2$. The $t$-and $u$-exchange amplitudes are projected to the S and P partial waves 
and these contributions are subtracted  from these amplitudes and added to the S- and P-wave in eqs.~(\ref{eq7}). The remaining $u$-and $t$-channel exchange amplitudes 
are essential to describe the high spin partial waves. 
\subsection{The observables}

The fit to the data yields the pole positions of resonances, the helicity amplitudes at the pole,
and the coupling constants for their decay to the different intermediate states. These quantities
are presented in the tables below. We also give the Breit-Wigner (BW) masses and widths. These are determined
by fitting a relativistic BW-amplitude 
to the pole position with the known couplings determined before as residues at the pole to the various intermediate states. 
In~\cite{Burkert:2022bqo}, different definitions of BRs were studied.
The conventional branching ratio is defined by
\begin{equation}
\Gamma_a(s) = \frac{g_a^2}{\sqrt{s_0}}\,\rho_a(s)\,n_a ^2(s)\,, \label{rel}
\end{equation}
where the coupling constants $g_a$ are parameters of the scattering amplitude used to fit the data.
The two-body phase space $\rho(s)$ is given by
\begin{equation}
\rho(s)=\frac{1}{16\pi}\frac{2p}{\sqrt{s}}\,,\nonumber
\end{equation}
where $p=\sqrt{(s-(m_1+m_2)^2)\,(s-(m_1-m_2)^2)}/(2\,\sqrt s)$,
and  the orbital-angular-momentum-barrier factor $n_a(s)$ for the resonance decay into channel $a$ by
\begin{equation}
n_a(s)= \left(\frac{p}{p_0}\right)^l \frac{F_l(z)}{F_l(z_0)}\,,
\end{equation}
with $z=(pR)^2$ and
\begin{align}
F_0^2= 1,  &\qquad F_1^2=\frac{1}{1+z}, &F_2^2=\frac{1}{9+3z+z^2}\,.
\end{align}
Higher order barrier factors can be found in Refs.~\cite{VonHippel:1972fg,Chung:1995dx}. 
The total width is given by the sum of all contributing partial widths.

Except for narrow resonances far above a threshold, BRs depend on the definition.
The best definition relies on an integration over the mass distribution of the produced resonance. In the
case of sequential decays, a double integration should be performed, taking into
account the mass distribution of the primarily produced resonance and the spectral function of the intermediate resonance.
In~\cite{Burkert:2022bqo}, the use of the formula
\begin{eqnarray}
\hspace{-7mm} \text{BR}_a =  \!\!\int\limits_{\rm threshold}^{\infty}\!\!\frac {ds}{\pi} \frac{ g_a^2\rho_{a}(s) n_a ^2(s)}{(M^2-s)^2+
(\sum\limits_i g_{i}^2\varrho_{i}(s) n_i ^2 (s))^2}, \label{br4}
\end{eqnarray}
was suggested. 
We are aware that this formula is an approximate one. 
Heuser {\it et al.}~\cite{Heuser:2024biq} have developed
a formalism consistent with the
fundamental principles of analyticity, unitarity, and positivity of the spectral function, allowing for a
precise definition of BRs via proper integrals over the given line shape. However, it can be used only
for two-body final states.

For the decay into $N\rho$, the phase volume
$\rho_f(s)=\rho_3(s)$ is given by the $N\pi\pi$ phase volume
\begin{eqnarray}
\hspace{-4mm} \rho_3(s)=\!\!\int\limits_{(m_1+m_2)^2}^{(Re\sqrt{s}-m_3)^2} \!\!\frac{ds_{r}}{\pi}
\frac{\rho(s,s_{r},L_{r},R_{r})\,M_{r}\,\Gamma_{r}(s_{r})} {(M_{r}^2\!-\!s_{r})^2\!+\!(M_{r}\,\Gamma_{r}(s_{r}))^2}\,,
\label{rho3}
\end{eqnarray}
where $M_{r}$, $\Gamma_{r}$, $L_{r}$, and $R_{r}$ are the mass, width,
orbital angular momentum and
range parameter of the intermediate resonance. The latter decays
into two particles with momenta $k_1$ and $k_2$, $s_{r}=(k_1+k_2)^2$, and
the relative momentum $k$.
The three-body phase space  in the quasi two-body approximation is given by
\begin{equation}
\rho(s,s_r,L_r,R_r)=\frac{1}{16\pi}\frac{2k}{\sqrt s}n_a(s)\,,
\end{equation}
and the width can be defined via the $P$-wave two-pion phase
space and the corresponding decay coupling $g_{\pi\pi}$
\begin{equation}
M_\rho\Gamma_{\rm \rho}=g_{\pi\pi}^2\rho(M^2,M^2_{\rho},L_\rho=1,R_\rho),
\end{equation}
with an effective radius $R_\rho$ of the $\rho$-meson.

We also give the $N\gamma$ partial decay width of resonances. They are calculated from 
\be
\Gamma_\gamma = \frac{k^2 _{BW}}{\pi}\frac{2m_N}{(2J+1)M_{BW}}\big( |A_{\nicefrac12}| ^2 + |A_{\nicefrac32}| ^2\big)\,,
\ee
where $k_{BW}$ is the decay momentum at the Breit-Wigner mass. 
\subsection{Fitting procedure}

For the data with two-body final states, the fit provides a $\chi^2$
per data point. The data with three-body final states are fitted event-by-event in a
likelihood fit. The fit minimizes
the pseudo-likelihood defined by
\begin{equation}
 -\ln {\cal L}_{\rm tot}= ( \frac 12\sum w_i\chi^2_i-\sum w_i\ln{\cal L}_i ) \ \frac{\sum
N_i}{\sum w_i N_i} \label{likeli}\,.
\end{equation}
The likelihood does not provide intuitive information
on the fit quality. Therefore, we calculate e.g. the $\chi^2$ per Dalitz plot cell which is given in Table~\ref{tab:useddata}.

To all datasets, a weight is assigned. When the weight of a new data set is increased, the likelihood contribution 
of the new data increases while $|\ln( {\cal L}_{\rm tot})|$ decreases. 
The weight is increased as long as the gain in $|\ln{\cal L}_i|$ for
the data set is larger than twice the loss in $|\ln({\cal L}_{\rm tot})|$.

The number of parameters is very large, and not all parameters can be left free in the fits at the same time.
When coupling constants were determined with uncertainties considerably larger than the assigned value, the
respective coupling constants were set to zero.

\subsection{Uncertainty evaluation} 
In order to evaluate the systematic uncertainties, we changed the weights of data sets and the model space as follows:
\begin{itemize}
\item The weight of the data on $\gamma p\to p\pi^+\pi^-$~\cite{CLAS:2018drk,CLAS:2024iir}
and $\vec\gamma \vec p\to p\pi^+\pi^-$~\cite{Crede:2024tbd} was changed by~a~factor~2~or~$\nicefrac12$. 
\item The weight of the data on $\gamma p\to p\pi^0\pi^0$~\cite{CBELSATAPS:2015kka}
and $\vec\gamma \vec p\to p\pi^0\pi^0$~\cite{CBELSATAPS:2022uad} was changed by a factor 2 or $\nicefrac12$. 
\item We either used only the mass- and angular distributions of the data in $\gamma p\to p\pi^+\pi^-$, or we used the event-based full data sample in~\cite{CLAS:2024iir}.
\item We either used the bubble-chamber on $\gamma p\to p\pi^+\pi^-$~~\cite{Aachen-Berlin-Bonn-Hamburg-Heidelberg-Munich:1968rzt}
to normalize the cross section or we did not.
\item Amplitudes for additional broad states at about 1900\,MeV in the $I(J^P)=\nicefrac12(\nicefrac32^+)$ ($N(1975)\nicefrac32^+$)
and $\nicefrac32(\nicefrac32^+)$ (hypothetical state) -waves were either included or not.
\item $N(2220)\nicefrac92^+$, or $N(2250)\nicefrac92^-$ were included in the fit or dismissed. 
\item The analyses were constrained by the real and imaginary part of the $\pi N$ elastic 
scattering amplitudes derived
alternatively by Koch~\cite{Koch:1985bp} or by the SAID group at GWU~\cite{Workman:2012hx}. 
The amplitudes were downloaded from the SAID data base~\cite{SAID} and used up to $W\leq 2.3$\,GeV.
\end{itemize}
For the KA84 amplitudes no uncertainties were evaluated by the author; 
we assume a 5\% relative uncertainty. Further, an absolute 0.005 
uncertainty (or 0.5\% of the unitarity limit $\Re A (\Im A)=1$) was added 
quadratically to the uncertainties in the KA84 and  WI08 amplitudes. 
Our uncertainties on masses, widths, and other properties are derived
from the spread of results when the fit model was changed.

\begin{table}[pt]
    \caption{Data on two-body final state used in this analysis and their $\chi^2$
    contribution to the final fits with KA84 or WI08 amplitudes. $\pi N\to \pi N$
    stands for the real and imaginary parts of the $\pi N$ scattering amplitudes
    in all partial waves with $L\leq 4$. }
    \label{tab:datatwobody}
    \centering{
    \renewcommand{\arraystretch}{1.35}
    \begin{tabular}{ccccc}
    \hline\hline
Reaction  &\phantom{zz} $N_{\rm data}$\phantom{zz} & $\chi^2/N_{\rm data, KA}$ &$\chi^2/N_{\rm data,WI08}$  \\\hline
$\gamma p\to N\pi$      &23615& 1.83&1.86\\
$\gamma p\to p\eta$     &4923 & 1.33&1.36\\
$\gamma p\to p\eta'$    &894  & 1.24&1.22\\
$\gamma p\to \Lambda K$ &4912 & 1.43&1.42\\    
$\gamma p\to \Sigma K$  &4543 & 1.51&1.54\\    
$\pi N\to \Lambda K$    &1092 & 1.52&1.51\\
$\pi N\to \Sigma K$     &1631 & 1.43&1.59\\
$\pi N\to \pi N$    &1534/1700 & 1.46& 2.12\\
\hline\hline
    \end{tabular}
        \caption{Binned data on two-pion production used in this analysis and their $\chi^2$
    contribution to the final fit. The $N_{\rm data}$
    is the number of bins. The $\chi^2$ contributions are given for
    the fit when KA84 amplitudes were used for $\pi N\to \pi N$. }
\begin{tabular}{ccccc}
    \hline\hline
Reaction  &~~~Obs.~~~&~~~$N_{\rm data}$~~~&~~~$\chi^2/N_{\rm data}$~~~&~Ref.~\\
\hline
$\gamma p\to p\pi^0\eta$   & $\Sigma$  & 90 & 1.61&\cite{CBELSATAPS:2014wvh}\\
$\gamma p\to p\pi^0\eta$   & $I^C;I^S$  & 60;60 & 1.67; 1.54 &\cite{CBELSATAPS:2014wvh}\\
$\gamma p\to p\pi^0\pi^0$   & $\Sigma$  & 121 & 2.48&\cite{CBELSATAPS:2015kka}\\
$\gamma p\to p\pi^0\pi^0$   & $H;P;T$  & 205;205;575 & 1.08;0.92;2.15 &\cite{CBELSATAPS:2022uad}\\
$\gamma p\to p\pi^0\pi^0$   & $P_x;P^c_x;P_x^s$  & 252;108;108 & 1.25;0.67;1.23 &\cite{CBELSATAPS:2022uad}\\
$\gamma p\to p\pi^0\pi^0$   & $P_y;P^c_y;P_y^s$  & 252;108;108 & 1.42;0.71;0.86 &\cite{CBELSATAPS:2022uad}\\
$\gamma p\to p\pi^0\pi^0$   & $I^C;I^S$  & 500;500 & 0.68; 1.86 &\cite{CBELSATAPS:2015tyg}\\
$\gamma p\to p\pi^+\pi^-$   & $I^C;I^S$  & 1200;1200 & 3.53;3.65 &\cite{CLAS:2025ogo}\\
$\gamma p\to p\pi^+\pi^-$   & $P_x;P^c_x;P_x^s$  & 1200;240;240 & 1.19;2.65;2.38 &\cite{CLAS:2025ogo}\\
$\gamma p\to p\pi^+\pi^-$   & $P_y;P^c_y;P_y^s$  & 1200;240;240 & 1.54;1.94;2.53 &\cite{CLAS:2025ogo}\\
    \hline\hline
    \end{tabular}
    \caption{Data on two-pion production used in this analysis. The $N_{\rm data}$
    is the number of cells in the Dalitz plots. The cells are not used in the fits;
    instead, the likelihood is maximized in an event-by-event fit. 
    The KA amplitudes were used for $\pi N\to \pi N$.
    }
    \label{tab:useddata}
    \centering
    \renewcommand{\arraystretch}{1.35}
\begin{tabular}{cccc}
    \hline\hline
Reaction  &\phantom{zz} $N_{\rm data}$\phantom{zz} & \phantom{zz}$\chi^2/N_{\rm data}$\phantom{zz}  &\phantom{zz} Ref.\phantom{zz} \\
\hline
$\gamma p\to p\pi^0\pi^0$ &665 ; 432; 648 & 1.17; 1.71; 0.89  & \cite{CBELSATAPS:2015kka}\\
$\gamma p\to p\pi^+\pi^-$ & 587; 558; 704 & 1.44; 1.22; 2.36& \cite{CLAS:2024iir}\\
$\pi^-p\to n\pi^0\pi^0$   & 217; 393 & 1.52; 1.76 & \cite{CrystalBall:2004qln}\\
$\pi^-p\to p\pi^+\pi^-$   & 1422; 1417 & 2.60; 2.57  & \cite{HADES:2020kce}\\
$\pi^-p\to p\pi^-\pi^0$   & 1090; 1211 & 2.30; 2.22  & \cite{HADES:2020kce}\\
    \hline\hline
    \end{tabular}
}
\end{table}

\section{\label{SectionResults}Results}

\subsection{Fit quality}
Table~\ref{tab:datatwobody} presents an overview of the data on single-meson production
used in the fits with the number of events and the $\chi^2$ per data point. 
The KA84 and WI08 amplitudes were used alternatively. In Table I, we
give the $\chi^2$ for fits constrained by either KA84 and or WI08.
For Tables II and III we give the result only for KA84 which defines our main
solution. The compatibility of the  KA84 amplitudes with the 
data on photoproduction of single mesons is somewhat better than of the WI08 amplitudes; 
overall, $\chi^2/N_{\rm data,KA}=1.73$ and $\chi^2/N_{\rm data,WI08}=1.84$. Most relevant
for this analysis are the data on two-pion production. For all data available event-based, these data are fit event-by-event, 
and only a likelihood is determined. In Table~\ref{tab:useddata} we give for the data shown in Fig.~\ref{fig:DP_CBELSA/TAPS} the number of cells in the Dalitz plot and $\chi^2$ per cell. The entries
per cell are not used in the fit. Of course,
the full data sets are used in the likelihood fit. Furthermore, the data on $\gamma p\to p\pi^0\pi^0$ from Refs.~\cite{Kashevarov:2012wy,Zehr:2012tj,Ajaka:2007zz,CBELSATAPS:2015tyg,CBELSATAPS:2022uad} (in part taken with polarized beam (and polarized target))
and on $\vec\gamma \vec p\to p\pi^+\pi^-$ from Ref.~\cite{Crede:2024tbd} are included in the
likelihood.

\subsection{Contributing resonances} 
\label{SectionResonances}
Table~\ref{tab:ND-mass} presents the list of 36 resonances used to describe the data. Masses and widths
are given at their pole position and in BW-approximation. The latter is determined by the procedure described in
Ref.~\cite{Anisovich:2011fc}.  

The agreement with the RPP values \cite{ParticleDataGroup:2024cfk} is, in general, very good. All masses are 
compatible with the RPP ranges; the widths of some resonances are found here wider than the RPP ranges. These are $N(1700)\nicefrac32^-$,$N(1875)\nicefrac32^-, N(1895)\nicefrac12^-, N(1900)\nicefrac32^+$, $N(2100)\nicefrac12^+$, and 
$\Delta(1900)\nicefrac12^-$. Results on masses and widths derived in Ref.~\cite{Mokeev:2023zhq}
are mostly compatible with the present findings; however, a few values agree only within $2\sigma$. Only 
the $N(1)700)\nicefrac32^-$ width reported here is considerably wider than in Ref.~\cite{Mokeev:2023zhq}
(and in RPP~\cite{ParticleDataGroup:2024cfk}).

Also given are the helicity amplitudes at the pole position and in BW-representation.
Our BW-helicity amplitudes agree well with the ranges given in the RPP. The values at the pole positions, in particular
the phases, determined by different groups are often inconsistent,
and the RPP does not give a range. 
When no range is given in the RPP, we refrain from comparing our results with previous determinations.
In some cases, the helicity amplitude has a phase compatible with 90$^\circ$; then, no sign is given
for the BW-helicity amplitude.

Table~\ref{yields} lists the decay BRs leading to the 
$\Delta\pi$, $N\rho$, and $Nf_0(500)$ 
final states. Here, large uncertainties
can be seen even for well-established resonances. 
Especially at higher masses,
results reported by different groups are often incompatible.

We determine BRs by integration over the width of a resonance and over the width of
an intermediate state~\cite{Burkert:2022bqo}. For high-mass resonances, the method how BRs
are determined has little influence on the result. For low-mass resonances, the method may have a significant
impact.

\subsection{Resonances in the second resonance region}

The $\Delta(1232)$ -- the only resonance in the first resonance region -- does not contribute to 
$\gamma p\to N\pi\pi$. The resonances 
\bc
\vspace{-3mm}\footnotesize
\begin{tabular}{ccc}
$N(1440)\nicefrac12^+$ & $N(1520)\nicefrac32^-$  & $N(1535)\nicefrac12^-$
\end{tabular}
\vspace{-3mm}
\ec
form the second resonance region. The lowest-mass excitation of the nucleon, $N(1440)\nicefrac12^+$, is often interpreted as a dynamically
generated resonance, but it is at least also the first radial excitation of the nucleon, as evidenced by the zero
in the electromagnetic helicity amplitude $A_{{1}/{2}}(Q^2)$~\cite{Eichmann:2018ytt}. The decay mode $Nf_0(500)$ 
of $N(1440)\nicefrac12^+$ observed here slightly exceeds 
the RPP range. Very recently, the pole position of this resonance has been determined
to  $M=$ (1374\er3\er4)\,MeV, $\Gamma=$ (215\er18\er8)\,MeV using constraints
from Roy-Steiner equations~\cite{Hoferichter:2023mgy}. Our result, $M=$ (1366\er3)\,MeV, $\Gamma=$ (192\er 4)\,MeV
fall slightly below the highly constrained determination of Ref.~\cite{Hoferichter:2023mgy}.

The two negative-parity states, $N(1520)\nicefrac32^-$ and $N(1535)\nicefrac12^-$, are determined with properties well consistent
with previous determinations, even though the $N\pi\pi$ couplings suffer from large uncertainties. Part of
the uncertainty can be traced to the $N(1520)\nicefrac32^-\to N\rho$ decays. The nominal mass of the $\rho$-meson 
plus the nucleon mass exceeds the $N(1520)\nicefrac32^-$ mass, hence the conventional BR vanishes. 
With integration over $N(1520)\nicefrac32^-$, we find a BR of (12\er2)\%, with double integration taking the width of the $\rho$ into account, we find 
(28\er4)\%.

\begin{table*}[thb]
    \caption{Masses and widths (in MeV) and helicity amplitudes (in GeV$^{-\nicefrac12}$) of
    $N^*$ and $\Delta^*$ resonances. The small numbers are the ranges given by the Particle Data Group. If PDG does not explicitly indicate a range, no PDG-numbers are given.
    For the Breit-Wigner helicity couplings marked with $^*$ only absolute values are given.}
    \label{tab:ND-mass}
    \footnotesize
    \centering
\renewcommand{\arraystretch}{1.4}
    \begin{tabular}{|r|ccccc|ccccc|cc|}
    \hline
  $N^*, \Delta^*$ $J^P$& $M_{\rm pole}$ & $\Gamma_{\rm pole}$ &   $M_{\rm BW}$ & $\Gamma_{\rm BW}$&&
 & $A_{\nicefrac12} ^{\rm pole}$ &phase& $A_{\nicefrac32}^{\rm pole}$ &phase& $A_{\nicefrac12} ^{\rm BW}$ & $A_{\nicefrac32}^{\rm BW}$  \\\hline
 $N(1440)$ $\nicefrac12 ^+$& 1366\er3       & 192\er4  & 1410\er10 & 290\er30 &&&
 -0.060\er0.006 & -(30\er7)\oo&&&-0.076\er0.008 &\\[-2ex]
                      &\tiny 1360-1380 &\tiny 180-205&\tiny 1410-1470&\tiny 250-450
                       &&&&&&&\tiny-0.065\er0.015&\\[-1ex]
$N(1520)$ $\nicefrac32 ^-$& 1506\er2 & 112\er3 & 1515\er3 & 114\er4 &&&
-0.028\er0.005 &  -(17\er8)\oo &   0.134\er0.007  &  (7\er4)\oo &-0.029\er0.004& 0.133\er0.006 \\[-2ex]
                      &\tiny 1505-1515 &\tiny 105-120&\tiny 1510-1520&\tiny 100-120
                       &&&&&&&\tiny-0.022\er0.008&\tiny0.140\er0.005\\[-1ex]
$N(1535)$ $\nicefrac12 ^-$& 1494\er7 & 115\er10 & 1517\er4 & 115\er7 &&&
0.110\er0.006 & (0\er5)\oo&&&0.102\er0.008&\\[-2ex]
                      &\tiny 1500-1520 &\tiny 80-130&\tiny 1515-1545&\tiny 125-175
                       &&&&&&&\tiny0.105\er0.015&\\[-1ex]
$N(1650)$ $\nicefrac12 ^-$& 1664\er10 & 105\er6 & 1670\er6 & 110\er7  &&&
0.031\er0.005&(2\er5)\oo&&&0.031\er0.004&\\[-2ex]
                      &\tiny 1650-1680 &\tiny 100-170&\tiny 1635-1665&\tiny 100-150
                       &&&&&&&\tiny0.045\er0.010&\\[-1ex]
$N(1675)$ $\nicefrac52 ^-$& 1662\er5 & 130\er8 & 1668\er4 & 130\er8  &&&
0.021\er0.004 &   30\er15\oo & 0.030\er0.004 &-(25\er10)\oo&0.022\er0.004 &0.030\er0.005\\[-2ex]
                      &\tiny 1650-1660 &\tiny 120-150&\tiny 1665-1680&\tiny 130-160
                       &&&&&&&\tiny0.018\er0.008&\tiny0.022\er0.008\\[-1ex]
$N(1680)$ $\nicefrac52 ^+$& 1676\er6 & 117\er4 & 1689\er5 & 122\er5  &&&
-0.016\er0.004&-(35\er15)\oo&  0.134\er0.005& (2\er5)\oo&-0.016\er0.004& 0.133\er0.005\\[-2ex]
                      &\tiny 1660-1680 &\tiny 110-135&\tiny {1680-1690}&\tiny 115-130
                       &&&&&&&\tiny-0.012\er0.007&\tiny0.135\er0.005\\[-1ex]
$N(1700)$ $\nicefrac32 ^-$& 1745\er40 & 420\er50 & 1770\er30 & 400\er50  &&&
 0.035\er0.014& -(20\er35)\oo &-0.100\er0.030&  (5\er20)\oo & 0.032\er0.013& -0.090\er0.025\\[-2ex]
                      &\tiny 1650-1750 &\tiny 100-300&\tiny 1650-1800&\tiny 100-300
                       &&&&&&&&\\[-1ex]
$N(1710)$ $\nicefrac12 ^+$& 1696\er10 & 155\er15 & 1710\er8 & 155\er14  &&&
0.045\er0.012 & -(30\er20)\oo & & & 0.049\er0.008 &\\[-2ex]
                      &\tiny 1650-1750 &\tiny 80-160&\tiny 1680-1740&\tiny 80-200
                       &&&&&&&&\\[-1ex]
$N(1720)$ $\nicefrac32 ^+$& 1710\er20 & 180\er30 & 1716\er18 & 180\er25  &&&
0.100\er0.025 &  -(5\er30)\oo & 0.045\er0.020 & (120\er30)\oo & 0.100\er0.025 & -0.041\er0.020\\[-2ex]
                      &\tiny 1660-1710 &\tiny 150-300&\tiny 1680-1750&\tiny 150-400
                       &&&&&&&\tiny0.100\er0.020&\\[-1ex]
$N(1860)$ $\nicefrac52 ^+$& 1870\er25 & 290\er40 & 1920\er25 & 300\er35  &&&
0.020\er0.006 &(85\er25)\oo & 0.062\er0.015 &  -(5\er15)\oo & 0.022\er0.006$^*$&  0.063\er0.013\\
$N(1875)$ $\nicefrac32 ^-$& 1855\er17 & 260\er20 & 1885\er15 & 270\er20  &&&
 0.012\er0.006 &  -(140\er30)\oo &  0.014\er0.006 & (175\er30)\oo &  -0.013\er0.006 &  -0.015\er0.006\\[-2ex]
                      &\tiny 1850-1950 &\tiny 100-220&\tiny 1850-1920&\tiny 120-250
                       &&&&&&&\tiny0.018\er0.008&\tiny-0.008\er0.018\\[-1ex]
$N(1880)$ $\nicefrac12 ^+$& 1860\er25 & 270\er30 & 1865\er25 & 275\er30  &&&
0.015\er0.005 & -(10\er20)\oo &&& 0.016\er0.005 &\\[-2ex]
                      &\tiny 1820-1900 &\tiny 180-280&\tiny 1830-1930&\tiny {200-400}
                       &&&&&&&&\\[-1ex]
$N(1895)$ $\nicefrac12 ^-$& 1905\er15 & 185\er25 & 1907\er15 & 195\er20  &&&
0.021\er0.006      &  (150\er30)\oo &&&  -0.021\er0.008& \\[-2ex]
                      &\tiny 1890-1930 &\tiny 80-140&\tiny {1870-1920}&\tiny {80-200}
                       &&&&&&&&\\[-1ex]
$N(1900)$ $\nicefrac32 ^+$& 1925\er25 & 270\er30 & 1930\er15 & 270\er20  &&&
 -0.050\er0.012&  (40\er40)\oo& -(0.044\er0.012)&  (35\er 20)\oo & -0.050\er0.012&-0.044\er0.014\\[-2ex]
                      &\tiny 1900-1940 &\tiny 90-160&\tiny 1890-1950&\tiny 100-320
                       &&&&&&&&\\[-1ex]
$N(1975)$ $\nicefrac32 ^+$& 1935\er60 & 640\er100 & 1950\er60 & 660\er100 &&&
0.075\er0.025&-(75\er40)\oo&{0.075\er0.030}&-(35\er40)\oo& {0.076\er0.025$^*$} &{0.077\er0.030}\\
$N(1990)$ $\nicefrac72 ^+$& 2005\er25 & 270\er30 & 2020\er25 & 280\er30  &&&
0.011\er0.006&(40\er20)\oo&0.017\er0.008&(90\er25)\oo&0.012\er0.010& 0.017\er0.008$^*$\\[-2ex]
                      &            &          &\tiny 1950-2100 &\tiny 200-400
                       &&&&&&&&\\[-1ex]
$N(2000)$ $\nicefrac52 ^+$& 1990\er40 & 480\er{60} & 2050\er40 & 500\er{60}  &&&
{0.033\er0.006}& {(35\er25)}\oo& \b{0.043\er 0.012}&{-(100\er30)\oo}& {0.033\er0.006}& {-0.044\er0.012}\\[-1ex]
$N(2060)$ $\nicefrac52 ^-$& 2020\er30 & 410\er{50} & 2040\er30 & 420\er50 &&&
0.065\er0.015& (20\er10)\oo&0.030\er0.020& (160\er30)\oo&0.065\er0.015 &-0.030\er0.020\\[-2ex]
                      &\tiny 2020-2130 &\tiny 350-430&\tiny 2030-2200&\tiny 300-450
                       &&&&&&&&\\[-1ex]
$N(2100)$ $\nicefrac12 ^+$& 2055\er25 & 430\er65 & 2070\er35 & 450\er60  &&&
 (0.020\er0.008)&   (65\er20)\oo & & & 0.021\er0.007&\\[-2ex]
                      &\tiny 2050-2150 &\tiny 240-340&\tiny 2050-2150&\tiny 200-320
                       &&&&&&&&\\
$N(2120)$ $\nicefrac32 ^-$& 2130\er40 & 350\er35 & 2140\er35 & 360\er35  &&&
-0.025\er0.015 &  (40\er25)\oo&-0.045\er0.015 &   (25\er25)\oo  &0.026\er0.015&0.047\er0.015\\[-2ex]
                      &\tiny 2050-2150 &\tiny 200-360&\tiny 2060-2160&\tiny 260-360
                       &&&&&&&&\\
$N(2190)$ $\nicefrac72 ^-$& 2130\er35 & 370\er45 & 2170\er25 & 420\er45 &&&
0.060\er0.012 &(150\er25)\oo &0.050\er0.015 & -(10\er25)\oo &-0.060\er0.010 &0.051\er0.006\\[-2ex]
                      &\tiny {1950}-2150 &\tiny 300-500&\tiny 2140-2220&\tiny 300-500
                       &&&&&&&&\\
$N(2220)$ $\nicefrac92 ^+$&2165\er30 & 440\er40& 2230\er30 &500\er40&&&0.015\er0.010 &(90\er30)\oo &0.018\er0.010 & -(20\er30)\oo & 0.017\er0.010$^*$&0.020\er0.010\\[-2ex]
                      &\tiny 2130-2200 &\tiny 360-480&\tiny 2200-2300&\tiny 350-500
                       &&&&&&&&\\
$N(2250)$ $\nicefrac92 ^-$&2220\er40 & 470\er40& 2310\er40 &  520\er 50&&&0.020\er0.010
 &-(10\er30)\oo & 0.025\er0.009& -(10\er20)\oo & 0.015\er0.008&0.025\er0.008\\[-2ex]
                      &\tiny 2100-2200 &\tiny 350-500&\tiny 2250-2320&\tiny 300-600
                       &&&&&&&&\\
    \hline
$\Delta(1232)$ $\nicefrac32 ^+$& 1210\er1       & 101\er2  & 1230\er1 & 116\er2&&
&-0.133\er0.003&-(17\er2)$^\circ$ &{-0.255\er0.005}&(-10\er2)$^\circ$
&-0.135\er0.003&-0.259\er0.005\\[-2ex]
                      &\tiny 1209-1211 &\tiny 98-102&\tiny 1230-1234&\tiny 114-120&
&\tiny &&&&
&\tiny -(0.129-0.142)&\tiny -(0.248-0.262)\\[-1ex]
$\Delta(1600)$ $\nicefrac32 ^+$& 1550\er15       & 260\er20  & 1575\er15 & {290\er25}&&
&0.040\er0.008&(120\er20)$^\circ$ &{0.038\er0.012}&(155\er20)$^\circ$
&-0.042\er0.010&-0.039\er0.008\\[-2ex]
                      &\tiny 1470-1590 &\tiny 150-320&\tiny 1500-1640&\tiny 200-300&
&\tiny &&&&
&\tiny -0.045\er 0.015&\tiny -0.035\er0.015\\[-1ex]
$\Delta(1620)$ $\nicefrac12 ^-$& 1598\er7 & 130\er10 & 1600\er8 & 135\er10&&
&0.047\er0.007&{(10\er15)}$^\circ$ &&&0.048\er0.007&\\[-2ex]
                      &\tiny 1590-1610 &\tiny 80-140&\tiny 1590-1630&\tiny 110-150
                       &&&&&&&\tiny 0.045\er 0.015&\\[-1ex]
$\Delta(1700)$ $\nicefrac32 ^-$& 1655\er15 & 300\er15 & 1690\er15 & 310\er15&&
&0.150\er0.020&(30\er15)$^\circ$ &0.155\er0.020&(40\er15)$^\circ$
& 0.153\er0.018&0.158\er0.017\\[-2ex]
                      &\tiny 1640-1690 &\tiny 200-300&\tiny 1690-1730&\tiny 220-380
                      &&&&&&&\tiny0.130\er0.030&\tiny0.130\er0.040\\[-1ex]
$\Delta(1750)$ $\nicefrac12 ^+$& 1770\er30 & 290\er45 & 1790\er30 & 305\er35  &&&
0.020\er0.007&-(85\er30)\oo&&&0.020\er0.007&\\
$\Delta(1900)$ $\nicefrac12 ^-$& 1815\er20 & 340\er35 & 1825\er20 & 360\er35&&
& 0.075\er0.015&-(65\er25)$^\circ$ && &0.075\er0.015&\\[-2ex]
                      &\tiny 1830-1900 &\tiny 180-320&\tiny 1840-1920&\tiny 180-{300}
                      &&&&&&&&\\[-1ex]
$\Delta(1905)$ $\nicefrac52 ^+$& 1795\er20 & 280\er20 & 1845\er20 & 315\er20&&
&0.030\er0.007&-(35\er15)$^\circ$ &-0.072\er0.010&-(10\er20)$^\circ$
&0.032\er0.008&-0.075\er0.010\\[-2ex]
                      &\tiny 1750-1800 &\tiny 260-340&\tiny 1855-1910&\tiny 270-400
                      &&&&&&&\tiny 0.022\er 0.005&\tiny -0.045\er 0.010\\[-1ex]
$\Delta(1910)$ $\nicefrac12 ^+$& {1880\er20} & {480\er65} & {1890\er20} & {520\er60}&&
& {0.090\er0.009}& {(40\er40)}$^\circ$ && & {0.100\er0.040}&\\[-2ex]
                      &\tiny 1800-1900 &\tiny 200-500&\tiny 1850-1950&\tiny 200-400
                      &&&&&&&\tiny 0.020\er 0.010&\\[-1ex]
$\Delta(1920)$ $\nicefrac32 ^+$& 1880\er30 & 260\er40 & 1890\er25 & 280\er40&&
&0.040\er0.025&-(30\er25)$^\circ$ &0.055\er0.025&-(50\er35)$^\circ$
& 0.043\er0.025&0.058\er0.030\\[-2ex]
                      &\tiny 1850-1950 &\tiny 200-400&\tiny 1870-1970&\tiny 240-360
                      &&&&&&&&\\[-1ex]
$\Delta(1930)$ $\nicefrac52 ^-$& 1812\er10 & 420\er25 & 1834\er10 & 425\er25  &&&
0.036\er0.008&  (117\er30)\oo& 0.020\er0.008& -(150\er30)\oo & -0.037\er 0.008& -0.020\er 0.008
\\[-2ex]
                      &\tiny 1820-1880 &\tiny 300-450&\tiny 1900-2000&\tiny 200-400
                      &&&&&&&&\\[-1ex]
$\Delta(1940)$ $\nicefrac32 ^-$   &2040\er40  &450\er50 & 2060\er40 & 460\er50 &&&
{0.045\er0.025}& {-(40\er30)\oo}& {0.080\er0.030}& {-(75\er30)}\oo&{0.052\er0.025}
&{0.095\er0.030}\\
$\Delta(1950)$ $\nicefrac72 ^+$& 1892\er5 & 250\er10 & 1919\er5 & 258\er8&&
&-0.076\er0.006&-(10\er5)$^\circ$ &-0.092\er0.004&-(10\er5)$^\circ$
&-0.077\er0.006&-0.094\er0.004\\[-2ex]
                      &\tiny 1870-1890 &\tiny 220-260&\tiny 1915-1950&\tiny 235-335
                      &&&&&&&\tiny-0.070\er0.005&\tiny-0.090\er0.010\\[-1ex]
$\Delta(2200)$ $\nicefrac72 ^-$&2090\er20& {360\er35} &{2180\er30} & {420\er50} &&&
{0.120\er0.015} & -(35\er20)\oo& 0.073\er{0.012} & -(30\er{20})\oo & {0.125\er 0.016}& {0.075\er 0.012}
\\[-2ex]
                      &\tiny 2050-2150 &\tiny 260-420&\tiny 2150-2250&\tiny 200-500&&
                      &&&&&&\\[-1ex]
$\Delta(2210)$ $\nicefrac52 ^-$&2200\er35&340\er40 &2205\er30 &345\er35&&
&0.057\er0.012&(7\er20)$^\circ$ &0.021\er0.010&(160\er25)$^\circ$
& 0.047\er0.012&-0.018\er0.010\\
     \hline\hline
   \end{tabular}
\renewcommand{\arraystretch}{1.0}
\end{table*}
\begin{table*}
\caption{\label{yields}
Branching ratios in \% for decays of $N^*$ and $\Delta^*$ resonances into $\Delta(1232)\pi$, $N\rho$, $Nf_0(500)$,
and the contribution of intermediate $N^*$ resonances to $N\pi\pi$. The small numbers are estimates of the Particle Data Group.
}
   \footnotesize
\renewcommand{\arraystretch}{1.4}
\begin{center}
\begin{tabular}{|rl|ccc|cccc|c|}
\hline\hline
                &&\multicolumn{1}{c}{ \phantom{z}$(\Delta\pi)_{\rm tot}$\phantom{z} }  &\multicolumn{2}{c|}{\phantom{zz} $\Delta\pi$\phantom{zz}}     &\multicolumn{1}{c}{\phantom{z} $(N\rho)_{\rm tot}$\phantom{z} } &\multicolumn{3}{c|}{\phantom{zzz}$N\rho$\phantom{zzz}}  &\phantom{zzzt}
$Nf_0(500)$\phantom{zzzt}\\\hline\\[-4.5ex]
                        &&&&&\multicolumn{1}{c}{ \phantom{zz} } &\multicolumn{1}{c}{\tiny\phantom{zz}$S=\nicefrac12$\phantom{zz}}  &{\tiny$S=\nicefrac32$}  &\multicolumn{1}{c|}{\tiny$S=\nicefrac32$}     &\\[-2ex]
                        &&&\multicolumn{1}{c}{\tiny \phantom{z}${L<J}$ \phantom{z}}& \multicolumn{1}{c|}{\tiny \phantom{z}${L>J}$ \phantom{z}}     &\multicolumn{1}{c}{ \phantom{zz} } & &\multicolumn{1}{c}{\phantom{zz}\tiny $L<J$\phantom{zz}}& \multicolumn{1}{c|}{\phantom{zz}\tiny $L>J$\phantom{zz}}     &  \\[-0.5ex]\hline
$N(1440)$&$\nicefrac12^+$                   &15\er7&-&15\er7&18\er6&9\er4&-&9\er4&15\er5    \\[-2ex]
&&\tiny{$$}&&\tiny{$6-27$}&\tiny{$$}&&&&\tiny{$11-13$}   \\[-1ex]
$N(1520)$&$\nicefrac32^-$                   &26\er6&12\er4&14\er4&28\er4&4\er3&24\er3&$<1$& $<2$    \\[-2ex]
&&\tiny{$22-34$}&\tiny{$15-{23}$}&\tiny{$7-11$} &\tiny{$10-16$}& \tiny{$0.2-0.4$} & \tiny{$10-16$}&\tiny{$$} & \\[-1ex]
$N(1535)$&$\nicefrac12^-$                    &5\er3&&5\er3&-&$<1$&&&6\er3\\[-2ex]
&& &&\tiny{$1-4$}  &\tiny{$2-17$} &\tiny{$2-16$}&& \tiny{$<1$} &\tiny{$2-10$} \\[-1ex]
$N(1650)$&$\nicefrac12^-$                    &6\er5&&6\er5&17\er6&12\er5&&5\er3&3\er2\\[-2ex]
&&\tiny{$$}&\tiny{$$}&\tiny{$6-18$}&\tiny{$12-22$}
                                  &\tiny{$<4$}&\tiny{$$}&\tiny{$12-18$}&\tiny{$2-18$}\\[-1ex]
$N(1675)$&$\nicefrac52^-$                     &19\er4&19\er4&&30\er9&20\er7&10\er5&&5\er3     \\[-2ex]
&&\tiny{$$}&\tiny{$23-37$}&&\tiny{$$}\tiny{$0.1-0.9$}
                                  &\tiny{$<0.2$}&\tiny{$0.1-0.7$}&\tiny{$$}&\tiny{$3-7$}\\[-1ex]
$N(1680)$&$\nicefrac52^+$                    &23\er7&9\er3&14\er6&9\er4&&9\er4&&      \\[-2ex]
&&\tiny{$11-23$}&\tiny{$4-10$}&\tiny{$1-13$}&\tiny{$8-11$}
                                  &\tiny{$$}&\tiny{$6-8$}&\tiny{$2-3$}&\tiny{$9-19$}\\[-1ex]
$N(1700)$&$\nicefrac32^-$                   &58\er8&54\er8&4\er2&21\er9&5\er3&-&16\er8&2\er2     \\[-2ex]
&&\tiny{$55-85$}&\tiny{$50-80$}&\tiny{$4-14$}&\tiny{$$}
                                  &\tiny{$$}&\tiny{$32-44$}&\tiny{$$}&\tiny{$2-14$}\\[-1ex]
$N(1710)$&$\nicefrac12^+$                     &7\er4&&7\er4&17\er4&6\er2&&11\er3&11\er4     \\[-2ex]
&&\tiny{$$}&\tiny{$$}&\tiny{$3-9$}&\tiny{$$}
                                  &\tiny{$11-23$}&\tiny{$$}&\tiny{$$}&\tiny{$<16$}\\[-1ex]
$N(1720)$&$\nicefrac32^+$                     &20\er7&9\er5&11\er5&29\er8&12\er5&17\er6&&25\er6       \\[-2ex]
&&\tiny{$47-89$}&\tiny{$47-77$}&\tiny{$<12$}&\tiny{$1-2$}
                                  &\tiny{$1-2$}&\tiny{$$}&\tiny{$$}&\tiny{$2-14$}\\[-1ex]
$N(1860)$&$\nicefrac52^+$                    &7\er3&2\er2&5\er2&54\er16&16\er9&13\er5&25\er12&9\er3     \\[-2ex]
&&\tiny{$20-54$}&\tiny{$4-16$}&\tiny{$16-38$}&\tiny{$<8.6$}
                                  &\tiny{$$}&\tiny{$<{8.5}$}&\tiny{$<0.1$}&\tiny{$41-61$}\\[-1ex]
$N(1875)$&$\nicefrac32^-$                      &14$^{+15}_{-10}$&10$^{+15}_{-10}$&4\er2&-&-&-&-&33\er20  \\[-1.5ex]
&&\tiny{$4-44$}&\tiny{$2-21$}&\tiny{$2-23$}&\tiny{$$}
                                  &\tiny{$$}&\tiny{$36-56$}&\tiny{$$}&\tiny{$16-60$}\\[-1ex]
$N(1880)$&$\nicefrac12^+$                     &6\er3&&6\er3&28\er7&20\er6&&8\er3&20\er4      \\[-2ex]
&&\tiny{$5-42$}&\tiny{$$}&&\tiny{$19-45$}&\tiny{$19-45$}
                                  &\tiny{$$}&\tiny{$$}&\tiny{$8-40$}\\[-1ex]
$N(1895)$&$\nicefrac12^-$                     &8\er3&&8\er3&43\er15&18\er5&&25\er14&18\er6     \\[-2ex]
&&\tiny{$$}&\tiny{$$}&\tiny{$3-11$}&\tiny{$14-50$}
                                  &\tiny{$<18$}&\tiny{$$}&\tiny{$14-32$}&\tiny{$<13$}\\[-1ex]
$N(1900)$&$\nicefrac32^+$                     &13\er5&8\er4&5\er3&46\er13&7\er4&9\er3&30\er12&10\er3     \\[-2ex]
&&\tiny{$30-70$}&\tiny{$9-25$}&\tiny{$21-45$}&\tiny{$$}
                                  &\tiny{$25-40$}&\tiny{$$}&\tiny{$$}&\tiny{$1-7$}
\\[-1ex]
$N(1975)$&$\nicefrac32^+$                     &15\er7&1\er1&14\er7&15\er5&1\er1&2\er2&12\er4&13\er5     \\
$N(1990)$&$\nicefrac72^+$                    &72\er15&72\er15&&10\er5&8\er4&2\er2&&    \\
$N(2000)$&$\nicefrac52^+$                    &25\er6&8\er4&17\er4&15\er4&8\er3&7\er3&&28\er10    \\[-2ex]
&&\tiny{$30-80$}&\tiny{$12-32$}&\tiny{$19-49$}&\tiny{$$}
                                  &\tiny{$$}&\tiny{$$}&\tiny{$$}&\tiny{$5-15$}\\[-1ex]
$N(2060)$&$\nicefrac52^-$                    &12\er3&12\er3&&28\er11&24\er10&4\er4&&3\er2    \\[-2ex]
&&\tiny{$$}&\tiny{$4-10$}&\tiny{$$}&\tiny{$5-33$}
                                  &\tiny{$<10$}&\tiny{$5-23$}&\tiny{$$}&\tiny{$3-9$}\\[-1ex]
$N(2100)$&$\nicefrac12^+$                    &10\er4&&10\er4&17\er7&12\er6&&5\er3&28\er6    \\[-2ex]
&&\tiny{$$}&\tiny{$$}&\tiny{$6-14$}&\tiny{$$}
                                  &\tiny{$35-70$}&\tiny{$$}&\tiny{$$}&\tiny{$14-35$}\\[-1ex]
$N(2120)$&$\nicefrac32^-$                     &19\er5&8\er3&11\er4&28\er6&4\er2&19\er5&5\er3&5\er3     \\[-2ex]
&&\tiny{$>23$}&\tiny{$15-70$}&\tiny{$8-45$}&\tiny{$$}
                                  &\tiny{$$}&\tiny{$<3$}&\tiny{$$}&\tiny{$4-15$}\\[-1ex]
$N(2190)$&$\nicefrac72^-$                    &4\er2&&&8\er7&&8\er7&     \\[-2ex]
&&\tiny{$$}&\tiny{$19-31$}&\tiny{$$}&\tiny{$$}
                            &\tiny{$$}&\tiny{$<11$}&\tiny{$$}&\tiny{$3-9$}\\[-1ex]
$N(2220)$&$\nicefrac92^+$     &15\er10& 15\er10& &            10\er10& & 10\er 10& & 5\er 5  \\
$N(2250)$&$\nicefrac92^-$  &10\er7&10\er7& & 11\er 7 & 5\er5& 6\er 4 & &  \\\hline

$\Delta(1600)$& $\nicefrac32^+$                  &44\er7&40\er6&4\er3&7\er4&2\er2&5\er3&&      \\[-2ex]
&&\tiny 58-82&\tiny 72-82&\tiny $<2$&\tiny&&&&   \\[-1ex]
$\Delta(1620)$&$\nicefrac12^-$                   &28\er11&&28\er11&52\er17&30\er12&&22\er12&     \\[-2ex]
&&&&\tiny44-72&\tiny23-32&\tiny23-32&&&  \\[-1ex]
$\Delta(1700)$&$\nicefrac32^-$                   &35\er13&23\er13&12\er8&15\er4&&15\er4&&     \\[-2ex]
&&\tiny 9-70 &\tiny 5-54&\tiny 4-16&&&\tiny 22-32&\tiny   &   \\[-1ex]
$\Delta(1750)$&$\nicefrac12^+$                   &32\er10&&32\er10&27\er13&17\er10&&10\er8&     \\
$\Delta(1900)$&$\nicefrac12^-$                    &64\er15&&64\er15&38\er13&18\er8&&20\er10&   \\[-2ex]
&&&&\tiny30-70&\tiny 22-60&\tiny11-35&&\tiny11-25&   \\[-1ex]
$\Delta(1905)$&$\nicefrac52^+$                     &19\er8&14\er7&5\er3&25\er10&&25\er10&&       \\[-2ex]
&&\tiny$>48$ &\tiny8-43&\tiny40-58 &\tiny &&\tiny 17-35&&  \\[-1ex]
$\Delta(1910)$&$\nicefrac12^+$                    &74\er10&&74\er10&10\er4&5\er3&&5\er3&   \\[-2ex]
&&\tiny34-66 &\tiny &&&\tiny&\tiny &&   \\[-1ex]
$\Delta(1920)$&$\nicefrac32^+$                    &38\er15&6\er4&32\er15&57\er8&8\er4&14\er5&35\er6&     \\[-2ex]
&&\tiny $>46$&\tiny 2-28&\tiny 44-72&&\tiny&\tiny &&      \\[-1ex]
$\Delta(1930)$&$\nicefrac52^-$                    &33\er9&28\er7&5\er5&33\er8&3\er2&&30\er8&      \\
$\Delta(1940)$&$\nicefrac32^-$                    &38\er10&21\er8&17\er9&24\er 6& 6\er4&10\er5&8\er4 & \\
$\Delta(1950)$&$\nicefrac72^+$                    &4\er3&4\er3&&10\er5&10\er5&&&    \\[-2ex]
&&\tiny &\tiny 1-9&&&\tiny&\tiny && \\[-1ex]
$\Delta(2200)$&$\nicefrac72^-$                     &2\er2&&1\er1&&&&&      \\[-2ex]
&&{\tiny$>$45} &{\tiny$>$40}   &{\tiny 5-25}   &&&&\tiny& \\
$\Delta(2210)$&$\nicefrac52^-$                    &40\er7&30\er6&10\er4&35\er8&&18\er6&17\er5 &     \\
\hline\hline
\end{tabular}
\end{center}
\end{table*}

\subsection{Resonances in the third resonance region}
Ten states observed can be assigned to the third resonance region: \\[-3ex]
\bc
\footnotesize
\begin{tabular}{ccc}
$N(1650)\nicefrac12^-$& $N(1700)\nicefrac32^-$& $N(1675)\nicefrac52^-$\\ 
$N(1710)\nicefrac12^+$& $N(1720)\nicefrac32^+$& $N(1680)\nicefrac52^+$\\ 
$\Delta(1750)\nicefrac12^+$&$\Delta(1600)\nicefrac32^+$&\\  
$\Delta(1620)\nicefrac12^-$ & $\Delta(1700)\nicefrac32^-$ \\[-1ex]
\end{tabular}
\ec
There is the well-known triplet of negative-parity 
$N^*$ states $N(1650)\nicefrac12^-$, $N(1700)\nicefrac32^-$, and $N(1675)\nicefrac52^-$ and the spin-doublet of $\Delta^*$'s,
$\Delta(1620)\nicefrac12^-$ and $\Delta(1700)\nicefrac32^-$. In the positive-parity sector, there seems to be a triplet, too.
However in quark models, $N(1720)\nicefrac32^+$ and $N(1680)\nicefrac52^+$ are interpreted as a spin doublet with total quark spin $S=\nicefrac12$ and intrinsic
orbital angular momentum $L=2$;  $N(1710)\nicefrac12^+$ is, like $N(1440)\nicefrac12^+$, a radial excitation of the
nucleon, but with a mixed-symmetry orbital excitation function, similar to that of $\Delta(1750)\nicefrac12^+$. 
For $N(1710)\nicefrac12^+$ and $\Delta(1750)\nicefrac12^+$ the spatial wave function includes both a radial and orbital excitation component. 

The partner of $N(1440)\nicefrac12^+$ with isospin $I=\nicefrac32$ is $\Delta(1600)\nicefrac32^+$, the radial excitation
of $\Delta(1232)$. Its mass reported in different analyses covers a wide range, 1470 to 1590\,MeV. 
Our mass value coincides with those of 
H\"ohler~\cite{Hohler:1993lbk} and Cutkosky et al.~\cite{Cutkosky:1980rh}, our width is compatible
with ~\cite{Cutkosky:1980rh}.

$\Delta(1600)\nicefrac32^+$ decays with a very high probability to
$\Delta(1232)\pi$; the RPP quotes 58\% to 82\%. We find (44\er7)\% by double integration. With integration
over the resonance only, our result of (67\er6)\% is just compatible with the RPP value.
For $\Delta(1620)\nicefrac12^-$, RPP reports 44\% to 72\% for 
$\Delta(1232)\pi$ decays; we find (28\er11)\% with double integration and (25\er10)\% with integration over $\Delta(1620)\nicefrac12^-$ only. For $N\rho$ decays, RPP reports (23-32)\%, we find (52\er17)\% with double and (28\er9)\% with single integration.
Large differences with different definition of branching ratios occur only for low-mass resonances.

Most other decay modes are seen with BRs compatible with RPP values. There are two exceptions: the
$N(1675)\nicefrac52^-\to N\rho$ BR is below 1\% in the only earlier determination~\cite{Hunt:2018wqz}
while we find a total of (30\er9)\%. $N(1720)\nicefrac32^+$ had a large BR for decays into $\Delta(1232)\pi$
(47-89)\%) and a small BR for $N\rho$ (1-2)\%, while we now find (20\er7)\% and (29\er8)\%, respectively.

\subsection{Resonances in the fourth and fifth resonance region}
The large number of resonances in the fourth 
\bc \vspace{-1mm}
\footnotesize
\begin{tabular}{cccc}
$N(1895)\nicefrac12^-$& $N(1875)\nicefrac32^-$& \\ 
$N(1880)\nicefrac12^+$& $N(1900)\nicefrac32^+$& $N(1860)\nicefrac52^+$& $N(1990)\nicefrac72^+$\\ 
              & $N(1975)\nicefrac32^+$& $N(2000)\nicefrac52^+$& \\ 
$\Delta(1900)\nicefrac12^-$& $\Delta(1940)\nicefrac32^-$&$\Delta(1930)\nicefrac52^-$& \\  
$\Delta(1910)\nicefrac12^+$& $\Delta(1920)\nicefrac32^+$&$\Delta(1905)\nicefrac52^+$& $\Delta(1950)\nicefrac72^+$\\  
\end{tabular}\\
\vspace{-1mm}
\ec
and fifth 
\bc \vspace{-1mm}
\footnotesize
\begin{tabular}{cccc}
& $N(2120)\nicefrac32^-$     & $N(2060)\nicefrac52^-$    & $N(2190)\nicefrac72^-$ \\ 
&$\Delta(2210)\nicefrac52^-$&$\Delta(2200)\nicefrac72^-$ &  \\  
& $N(2100)\nicefrac12^+$   & $N(2220)\nicefrac92^+$    & $N(2250)\nicefrac92^-$ \\
\end{tabular} \vspace{-1mm}
\ec
resonance region makes the assignment of contributions to the intermediate $ \Delta(1232)\nicefrac32^- \pi$ or $N\rho$-states difficult. 
Indeed, the discrepancies between different analyses become larger and larger with increasing resonance
mass. For many high-mass resonances, several decay modes are reported here for the first time.
In fits with very large numbers of free parameters, a multitude of different solutions exist, often
with very similar likelihood. We have performed a large number of tests to find the optimum fit and 
to determine realistic uncertainties. Still, the detailed numbers need to be taken with some 
caution.

\section{Summary}
\label{SectionSummary}
We have performed a coupled-channel analysis of data on two-pion production in pion- and photo-induced reactions
from BNL, ELSA, JLab, GSI, and MAMI. In photoproduction, the data include recent measurements where polarization
observables were determined. The fits are constrained by a large number of data on single-meson
production also including polarization observables.

The fit returned properties of 36 $N^*$ and $\Delta^*$ resonances: masses, widths and helicity amplitudes 
are determined at the pole position of resonances and in Breit-Wigner approximation.
Branching ratios are given
for $N^*$ and $\Delta^*$ resonances decaying into $\Delta(1232)\pi$, $N\rho(770)$, and $Nf_0(500)$ as intermediate
states. Decays via higher-mass $N^*$ resonances like $N(1440)\nicefrac12^+$, $N(1520)\nicefrac32^-$, $N(1535)\nicefrac12^-$, $N(1680)\nicefrac52^+\pi$, 
$N(1720)\nicefrac32^+\pi$ are admitted. The results substantially increase our knowledge on the decay of light-quark
baryon resonances.  We enphasize that four of these resonances
$N(1975)\nicefrac{3}{2}^+$, $\Delta(1750)\nicefrac{1}{2}^+$,  $\Delta(2190)\nicefrac{3}{2}^-$, and
 $\Delta(2210)\nicefrac{5}{2}^-$, improve the quality of the fit but we do not yet claim that these findings
 represent real resonances. Further detailed studies are required to understand why the fit including
 these additional resonances improves.\\[2ex]

{\small
The data on two-body final states used in this analyses are available on the BnGa web page (see Ref.~\cite{BnGa-web}).
The event-based data on three-body final states are available on request from the authors.\\[2ex]

The work was supported by the \textit{Deutsche Forschungsgemeinschaft} (through the funds provided to the Sino-German Collaborative Research Center TRR110 “Symmetries and the Emergence of Structure in QCD” (NSFC Grant No. 12070131001, DFG Project-ID 196253076 - TRR 110), by the programme “Netzwerke 2021”, 
an initiative of the Ministry of Culture and Science of the State of Northrhine Westphalia (project “NRW-FAIR”, ID: NW21-024-C), 
the EU Horizon 2020 research and innovation program, STRONG-2020 (grant agreement No. 824093), and 
the \textit{U.S. Department of Energy} (DE-AC05-06OR23177). Volker Crede acknowledges support from the U.S. Department of Energy, Office of Science, Office of Nuclear Physics, under Contract No. DE-FG02-92ER40735.}
\bibliography{Baryon}

\begin{thebibliography}{98}%
\makeatletter
\providecommand \@ifxundefined [1]{%
 \@ifx{#1\undefined}
}%
\providecommand \@ifnum [1]{%
 \ifnum #1\expandafter \@firstoftwo
 \else \expandafter \@secondoftwo
 \fi
}%
\providecommand \@ifx [1]{%
 \ifx #1\expandafter \@firstoftwo
 \else \expandafter \@secondoftwo
 \fi
}%
\providecommand \natexlab [1]{#1}%
\providecommand \enquote  [1]{``#1''}%
\providecommand \bibnamefont  [1]{#1}%
\providecommand \bibfnamefont [1]{#1}%
\providecommand \citenamefont [1]{#1}%
\providecommand \href@noop [0]{\@secondoftwo}%
\providecommand \href [0]{\begingroup \@sanitize@url \@href}%
\providecommand \@href[1]{\@@startlink{#1}\@@href}%
\providecommand \@@href[1]{\endgroup#1\@@endlink}%
\providecommand \@sanitize@url [0]{\catcode `\\12\catcode `\$12\catcode `\&12\catcode `\#12\catcode `\^12\catcode `\_12\catcode `\%12\relax}%
\providecommand \@@startlink[1]{}%
\providecommand \@@endlink[0]{}%
\providecommand \url  [0]{\begingroup\@sanitize@url \@url }%
\providecommand \@url [1]{\endgroup\@href {#1}{\urlprefix }}%
\providecommand \urlprefix  [0]{URL }%
\providecommand \Eprint [0]{\href }%
\providecommand \doibase [0]{https://doi.org/}%
\providecommand \selectlanguage [0]{\@gobble}%
\providecommand \bibinfo  [0]{\@secondoftwo}%
\providecommand \bibfield  [0]{\@secondoftwo}%
\providecommand \translation [1]{[#1]}%
\providecommand \BibitemOpen [0]{}%
\providecommand \bibitemStop [0]{}%
\providecommand \bibitemNoStop [0]{.\EOS\space}%
\providecommand \EOS [0]{\spacefactor3000\relax}%
\providecommand \BibitemShut  [1]{\csname bibitem#1\endcsname}%
\let\auto@bib@innerbib\@empty
\bibitem [{\citenamefont {Anisovich}\ \emph {et~al.}(2012)\citenamefont {Anisovich} \emph {et~al.}}]{Anisovich:2011fc}%
  \BibitemOpen
  \bibfield  {author} {\bibinfo {author} {\bibfnamefont {A.~V.}\ \bibnamefont {Anisovich}} \emph {et~al.},\ }\bibfield  {title} {\bibinfo {title} {{Properties of baryon resonances from a multichannel partial wave analysis}},\ }\href {https://doi.org/10.1140/epja/i2012-12015-8} {\bibfield  {journal} {\bibinfo  {journal} {Eur. Phys. J. A}\ }\textbf {\bibinfo {volume} {48}},\ \bibinfo {pages} {15} (\bibinfo {year} {2012})},\ \Eprint {https://arxiv.org/abs/1112.4937} {arXiv:1112.4937 [hep-ph]} \BibitemShut {NoStop}%
\bibitem [{\citenamefont {Gutz}\ \emph {et~al.}(2014)\citenamefont {Gutz} \emph {et~al.}}]{CBELSATAPS:2014wvh}%
  \BibitemOpen
  \bibfield  {author} {\bibinfo {author} {\bibfnamefont {E.}~\bibnamefont {Gutz}} \emph {et~al.} (\bibinfo {collaboration} {CBELSA/TAPS}),\ }\bibfield  {title} {\bibinfo {title} {{High statistics study of the reaction $\gamma p\to p\pi^0\eta$}},\ }\href {https://doi.org/10.1140/epja/i2014-14074-1} {\bibfield  {journal} {\bibinfo  {journal} {Eur. Phys. J. A}\ }\textbf {\bibinfo {volume} {50}},\ \bibinfo {pages} {74} (\bibinfo {year} {2014})},\ \Eprint {https://arxiv.org/abs/1402.4125} {arXiv:1402.4125 [nucl-ex]} \BibitemShut {NoStop}%
\bibitem [{\citenamefont {Sokhoyan}\ \emph {et~al.}(2015{\natexlab{a}})\citenamefont {Sokhoyan} \emph {et~al.}}]{CBELSATAPS:2015kka}%
  \BibitemOpen
  \bibfield  {author} {\bibinfo {author} {\bibfnamefont {V.}~\bibnamefont {Sokhoyan}} \emph {et~al.} (\bibinfo {collaboration} {CBELSA/TAPS}),\ }\bibfield  {title} {\bibinfo {title} {{High-statistics study of the reaction $\gamma p\to p\;2\pi^0$}},\ }\href {https://doi.org/10.1140/epja/i2015-15095-x} {\bibfield  {journal} {\bibinfo  {journal} {Eur. Phys. J. A}\ }\textbf {\bibinfo {volume} {51}},\ \bibinfo {pages} {95} (\bibinfo {year} {2015}{\natexlab{a}})},\ \bibinfo {note} {[Erratum: Eur.Phys.J.A 51, 187 (2015)]},\ \Eprint {https://arxiv.org/abs/1507.02488} {arXiv:1507.02488 [nucl-ex]} \BibitemShut {NoStop}%
\bibitem [{\citenamefont {Hunt}\ and\ \citenamefont {Manley}(2019)}]{Hunt:2018wqz}%
  \BibitemOpen
  \bibfield  {author} {\bibinfo {author} {\bibfnamefont {B.~C.}\ \bibnamefont {Hunt}}\ and\ \bibinfo {author} {\bibfnamefont {D.~M.}\ \bibnamefont {Manley}},\ }\bibfield  {title} {\bibinfo {title} {{Updated determination of $N^*$ resonance parameters using a unitary, multichannel formalism}},\ }\href {https://doi.org/10.1103/PhysRevC.99.055205} {\bibfield  {journal} {\bibinfo  {journal} {Phys. Rev. C}\ }\textbf {\bibinfo {volume} {99}},\ \bibinfo {pages} {055205} (\bibinfo {year} {2019})},\ \Eprint {https://arxiv.org/abs/1810.13086} {arXiv:1810.13086 [nucl-ex]} \BibitemShut {NoStop}%
\bibitem [{\citenamefont {M\"uller}\ \emph {et~al.}(2020)\citenamefont {M\"uller} \emph {et~al.}}]{CBELSATAPS:2019ylw}%
  \BibitemOpen
  \bibfield  {author} {\bibinfo {author} {\bibfnamefont {J.}~\bibnamefont {M\"uller}} \emph {et~al.} (\bibinfo {collaboration} {CBELSA/TAPS}),\ }\bibfield  {title} {\bibinfo {title} {{New data on $\vec{\gamma} \vec{p}\rightarrow \eta p$ with polarized photons and protons and their implications for $N^* \to N\eta$ decays}},\ }\href {https://doi.org/10.1016/j.physletb.2020.135323} {\bibfield  {journal} {\bibinfo  {journal} {Phys. Lett. B}\ }\textbf {\bibinfo {volume} {803}},\ \bibinfo {pages} {135323} (\bibinfo {year} {2020})},\ \Eprint {https://arxiv.org/abs/1909.08464} {arXiv:1909.08464 [nucl-ex]} \BibitemShut {NoStop}%
\bibitem [{\citenamefont {Afzal}\ \emph {et~al.}(2020)\citenamefont {Afzal} \emph {et~al.}}]{CBELSATAPS:2020cwk}%
  \BibitemOpen
  \bibfield  {author} {\bibinfo {author} {\bibfnamefont {F.}~\bibnamefont {Afzal}} \emph {et~al.} (\bibinfo {collaboration} {CBELSA/TAPS}),\ }\bibfield  {title} {\bibinfo {title} {{Observation of the $p\eta$' Cusp in the New Precise Beam Asymmetry $\Sigma$ Data for $\gamma p\to p \eta$}},\ }\href {https://doi.org/10.1103/PhysRevLett.125.152002} {\bibfield  {journal} {\bibinfo  {journal} {Phys. Rev. Lett.}\ }\textbf {\bibinfo {volume} {125}},\ \bibinfo {pages} {152002} (\bibinfo {year} {2020})},\ \Eprint {https://arxiv.org/abs/2009.06248} {arXiv:2009.06248 [nucl-ex]} \BibitemShut {NoStop}%
\bibitem [{\citenamefont {R\"onchen}\ \emph {et~al.}(2022)\citenamefont {R\"onchen}, \citenamefont {D\"oring}, \citenamefont {Mei\ss{}ner},\ and\ \citenamefont {Shen}}]{Ronchen:2022hqk}%
  \BibitemOpen
  \bibfield  {author} {\bibinfo {author} {\bibfnamefont {D.}~\bibnamefont {R\"onchen}}, \bibinfo {author} {\bibfnamefont {M.}~\bibnamefont {D\"oring}}, \bibinfo {author} {\bibfnamefont {U.-G.}\ \bibnamefont {Mei\ss{}ner}},\ and\ \bibinfo {author} {\bibfnamefont {C.-W.}\ \bibnamefont {Shen}},\ }\bibfield  {title} {\bibinfo {title} {{Light baryon resonances from a coupled-channel study including $\mathbf{K\Sigma}$ photoproduction}},\ }\href {https://doi.org/10.1140/epja/s10050-022-00852-1} {\bibfield  {journal} {\bibinfo  {journal} {Eur. Phys. J. A}\ }\textbf {\bibinfo {volume} {58}},\ \bibinfo {pages} {229} (\bibinfo {year} {2022})},\ \Eprint {https://arxiv.org/abs/2208.00089} {arXiv:2208.00089 [nucl-th]} \BibitemShut {NoStop}%
\bibitem [{\citenamefont {Navas}\ \emph {et~al.}(2024)\citenamefont {Navas} \emph {et~al.}}]{ParticleDataGroup:2024cfk}%
  \BibitemOpen
  \bibfield  {author} {\bibinfo {author} {\bibfnamefont {S.}~\bibnamefont {Navas}} \emph {et~al.} (\bibinfo {collaboration} {Particle Data Group}),\ }\bibfield  {title} {\bibinfo {title} {{Review of particle physics}},\ }\href {https://doi.org/10.1103/PhysRevD.110.030001} {\bibfield  {journal} {\bibinfo  {journal} {Phys. Rev. D}\ }\textbf {\bibinfo {volume} {110}},\ \bibinfo {pages} {030001} (\bibinfo {year} {2024})}\BibitemShut {NoStop}%
\bibitem [{\citenamefont {Klempt}\ and\ \citenamefont {Richard}(2010)}]{Klempt:2009pi}%
  \BibitemOpen
  \bibfield  {author} {\bibinfo {author} {\bibfnamefont {E.}~\bibnamefont {Klempt}}\ and\ \bibinfo {author} {\bibfnamefont {J.-M.}\ \bibnamefont {Richard}},\ }\bibfield  {title} {\bibinfo {title} {{Baryon spectroscopy}},\ }\href {https://doi.org/10.1103/RevModPhys.82.1095} {\bibfield  {journal} {\bibinfo  {journal} {Rev. Mod. Phys.}\ }\textbf {\bibinfo {volume} {82}},\ \bibinfo {pages} {1095} (\bibinfo {year} {2010})},\ \Eprint {https://arxiv.org/abs/0901.2055} {arXiv:0901.2055 [hep-ph]} \BibitemShut {NoStop}%
\bibitem [{\citenamefont {Crede}\ and\ \citenamefont {Roberts}(2013)}]{Crede:2013kia}%
  \BibitemOpen
  \bibfield  {author} {\bibinfo {author} {\bibfnamefont {V.}~\bibnamefont {Crede}}\ and\ \bibinfo {author} {\bibfnamefont {W.}~\bibnamefont {Roberts}},\ }\bibfield  {title} {\bibinfo {title} {{Progress towards understanding baryon resonances}},\ }\href {https://doi.org/10.1088/0034-4885/76/7/076301} {\bibfield  {journal} {\bibinfo  {journal} {Rept. Prog. Phys.}\ }\textbf {\bibinfo {volume} {76}},\ \bibinfo {pages} {076301} (\bibinfo {year} {2013})},\ \Eprint {https://arxiv.org/abs/1302.7299} {arXiv:1302.7299 [nucl-ex]} \BibitemShut {NoStop}%
\bibitem [{\citenamefont {Thiel}\ \emph {et~al.}(2022)\citenamefont {Thiel}, \citenamefont {Afzal},\ and\ \citenamefont {Wunderlich}}]{Thiel:2022xtb}%
  \BibitemOpen
  \bibfield  {author} {\bibinfo {author} {\bibfnamefont {A.}~\bibnamefont {Thiel}}, \bibinfo {author} {\bibfnamefont {F.}~\bibnamefont {Afzal}},\ and\ \bibinfo {author} {\bibfnamefont {Y.}~\bibnamefont {Wunderlich}},\ }\bibfield  {title} {\bibinfo {title} {{Light Baryon Spectroscopy}},\ }\href {https://doi.org/10.1016/j.ppnp.2022.103949} {\bibfield  {journal} {\bibinfo  {journal} {Prog. Part. Nucl. Phys.}\ }\textbf {\bibinfo {volume} {125}},\ \bibinfo {pages} {103949} (\bibinfo {year} {2022})},\ \Eprint {https://arxiv.org/abs/2202.05055} {arXiv:2202.05055 [nucl-ex]} \BibitemShut {NoStop}%
\bibitem [{\citenamefont {Burkert}\ \emph {et~al.}(2022)\citenamefont {Burkert}, \citenamefont {Klempt},\ and\ \citenamefont {Thoma}}]{Burkert:2022adb}%
  \BibitemOpen
  \bibfield  {author} {\bibinfo {author} {\bibfnamefont {V.}~\bibnamefont {Burkert}}, \bibinfo {author} {\bibfnamefont {E.}~\bibnamefont {Klempt}},\ and\ \bibinfo {author} {\bibfnamefont {U.}~\bibnamefont {Thoma}},\ }\href@noop {} {\bibinfo {title} {{Light-quark baryons}}} (\bibinfo {year} {2022}),\ \Eprint {https://arxiv.org/abs/2211.12906} {arXiv:2211.12906 [hep-ph]} \BibitemShut {NoStop}%
\bibitem [{\citenamefont {Gross}\ \emph {et~al.}(2023)\citenamefont {Gross}, \citenamefont {Klempt} \emph {et~al.}}]{Gross:2022hyw}%
  \BibitemOpen
  \bibfield  {author} {\bibinfo {author} {\bibfnamefont {F.}~\bibnamefont {Gross}}, \bibinfo {author} {\bibfnamefont {E.}~\bibnamefont {Klempt}}, \emph {et~al.},\ }\bibfield  {title} {\bibinfo {title} {{50 Years of Quantum Chromodynamics}},\ }\href@noop {} {\bibfield  {journal} {\bibinfo  {journal} {Eur. Phys. J. C}\ }\textbf {\bibinfo {volume} {83}},\ \bibinfo {pages} {1} (\bibinfo {year} {2023})},\ \Eprint {https://arxiv.org/abs/2212.11107} {arXiv:2212.11107 [hep-ph]} \BibitemShut {NoStop}%
\bibitem [{\citenamefont {Seifen}\ \emph {et~al.}(2022)\citenamefont {Seifen} \emph {et~al.}}]{CBELSATAPS:2022uad}%
  \BibitemOpen
  \bibfield  {author} {\bibinfo {author} {\bibfnamefont {T.}~\bibnamefont {Seifen}} \emph {et~al.} (\bibinfo {collaboration} {CBELSA/TAPS}),\ }\href@noop {} {\bibinfo {title} {{Polarization observables in double neutral pion photoproduction}}} (\bibinfo {year} {2022}),\ \Eprint {https://arxiv.org/abs/2207.01981} {arXiv:2207.01981 [nucl-ex]} \BibitemShut {NoStop}%
\bibitem [{\citenamefont {Golovatch}\ \emph {et~al.}(2019)\citenamefont {Golovatch} \emph {et~al.}}]{CLAS:2018drk}%
  \BibitemOpen
  \bibfield  {author} {\bibinfo {author} {\bibfnamefont {E.}~\bibnamefont {Golovatch}} \emph {et~al.} (\bibinfo {collaboration} {CLAS}),\ }\bibfield  {title} {\bibinfo {title} {{First results on nucleon resonance photocouplings from the $\gamma p \to \pi^+\pi^-p$ reaction}},\ }\href {https://doi.org/10.1016/j.physletb.2018.10.013} {\bibfield  {journal} {\bibinfo  {journal} {Phys. Lett. B}\ }\textbf {\bibinfo {volume} {788}},\ \bibinfo {pages} {371} (\bibinfo {year} {2019})},\ \Eprint {https://arxiv.org/abs/1806.01767} {arXiv:1806.01767 [nucl-ex]} \BibitemShut {NoStop}%
\bibitem [{\citenamefont {Sarantsev}\ \emph {et~al.}(2024)\citenamefont {Sarantsev} \emph {et~al.}}]{CLAS:2024iir}%
  \BibitemOpen
  \bibfield  {author} {\bibinfo {author} {\bibfnamefont {A.~V.}\ \bibnamefont {Sarantsev}} \emph {et~al.} (\bibinfo {collaboration} {CLAS}),\ }\href@noop {} {\bibinfo {title} {{Photoproduction of two charged pions off protons in the resonance region}}} (\bibinfo {year} {2024}),\ \Eprint {https://arxiv.org/abs/2411.15423} {arXiv:2411.15423 [nucl-ex]} \BibitemShut {NoStop}%
\bibitem [{\citenamefont {Roy}\ \emph {et~al.}(2024)\citenamefont {Roy} \emph {et~al.}}]{Crede:2024tbd}%
  \BibitemOpen
  \bibfield  {author} {\bibinfo {author} {\bibfnamefont {P.}~\bibnamefont {Roy}} \emph {et~al.},\ }\href@noop {} {} (\bibinfo {year} {2024}),\ \bibinfo {note} {{``Measurement of (double-) polarization observables in the photoproduction of $\pi^+\pi^-$~meson pairs off the proton using CLAS at Jefferson Laboratory", in preparation}}\BibitemShut {NoStop}%
\bibitem [{\citenamefont {Manley}\ \emph {et~al.}(1984)\citenamefont {Manley}, \citenamefont {Arndt}, \citenamefont {Goradia},\ and\ \citenamefont {Teplitz}}]{Manley:1984jz}%
  \BibitemOpen
  \bibfield  {author} {\bibinfo {author} {\bibfnamefont {D.~M.}\ \bibnamefont {Manley}}, \bibinfo {author} {\bibfnamefont {R.~A.}\ \bibnamefont {Arndt}}, \bibinfo {author} {\bibfnamefont {Y.}~\bibnamefont {Goradia}},\ and\ \bibinfo {author} {\bibfnamefont {V.~L.}\ \bibnamefont {Teplitz}},\ }\bibfield  {title} {\bibinfo {title} {{An isobar model partial wave analysis of $\pi N \to\pi\pi N$ in the center-of-mass energy range 1320\,MeV to 1930\,MeV}},\ }\href {https://doi.org/10.1103/PhysRevD.30.904} {\bibfield  {journal} {\bibinfo  {journal} {Phys. Rev.}\ }\textbf {\bibinfo {volume} {D30}},\ \bibinfo {pages} {904} (\bibinfo {year} {1984})}\BibitemShut {NoStop}%
\bibitem [{\citenamefont {Longacre}\ \emph {et~al.}(1975)\citenamefont {Longacre} \emph {et~al.}}]{Longacre:1974xu}%
  \BibitemOpen
  \bibfield  {author} {\bibinfo {author} {\bibfnamefont {R.}~\bibnamefont {Longacre}} \emph {et~al.},\ }\bibfield  {title} {\bibinfo {title} {{$N^*$ Resonance Parameters and $K$-Matrix Fits to the Reactions $\pi + N \to \Delta \pi + \rho N + \epsilon n$}},\ }\href {https://doi.org/10.1016/0370-2693(75)90373-1} {\bibfield  {journal} {\bibinfo  {journal} {Phys. Lett. B}\ }\textbf {\bibinfo {volume} {55}},\ \bibinfo {pages} {415} (\bibinfo {year} {1975})}\BibitemShut {NoStop}%
\bibitem [{\citenamefont {Longacre}\ \emph {et~al.}(1978)\citenamefont {Longacre} \emph {et~al.}}]{Longacre:1977ga}%
  \BibitemOpen
  \bibfield  {author} {\bibinfo {author} {\bibfnamefont {R.~S.}\ \bibnamefont {Longacre}} \emph {et~al.},\ }\bibfield  {title} {\bibinfo {title} {{K-Matrix Fits to $\pi n \to n \pi$ and $\pi n \to n \pi \pi$ in the Resonance Region $\sqrt s$ = 1.3\,GeV to 2.0\,GeV}},\ }\href {https://doi.org/10.1103/PhysRevD.17.1795} {\bibfield  {journal} {\bibinfo  {journal} {Phys. Rev. D}\ }\textbf {\bibinfo {volume} {17}},\ \bibinfo {pages} {1795} (\bibinfo {year} {1978})}\BibitemShut {NoStop}%
\bibitem [{\citenamefont {Barnham}\ \emph {et~al.}(1980)\citenamefont {Barnham} \emph {et~al.}}]{Barnham:1980za}%
  \BibitemOpen
  \bibfield  {author} {\bibinfo {author} {\bibfnamefont {K.~W.~J.}\ \bibnamefont {Barnham}} \emph {et~al.},\ }\bibfield  {title} {\bibinfo {title} {{An Isobar Model Partial Wave Analysis of Three-body Final States in $\pi^+ p$ Interactions From Threshold to 1700\,MeV Center-of-mass Energy}},\ }\href {https://doi.org/10.1016/0550-3213(80)90109-1} {\bibfield  {journal} {\bibinfo  {journal} {Nucl. Phys. B}\ }\textbf {\bibinfo {volume} {168}},\ \bibinfo {pages} {243} (\bibinfo {year} {1980})}\BibitemShut {NoStop}%
\bibitem [{\citenamefont {Vrana}\ \emph {et~al.}(2000)\citenamefont {Vrana}, \citenamefont {Dytman},\ and\ \citenamefont {Lee}}]{Vrana:1999nt}%
  \BibitemOpen
  \bibfield  {author} {\bibinfo {author} {\bibfnamefont {T.~P.}\ \bibnamefont {Vrana}}, \bibinfo {author} {\bibfnamefont {S.~A.}\ \bibnamefont {Dytman}},\ and\ \bibinfo {author} {\bibfnamefont {T.~S.~H.}\ \bibnamefont {Lee}},\ }\bibfield  {title} {\bibinfo {title} {{Baryon resonance extraction from $\pi N$ data using a unitary multichannel model}},\ }\href {https://doi.org/10.1016/S0370-1573(99)00108-8} {\bibfield  {journal} {\bibinfo  {journal} {Phys. Rept.}\ }\textbf {\bibinfo {volume} {328}},\ \bibinfo {pages} {181} (\bibinfo {year} {2000})},\ \Eprint {https://arxiv.org/abs/nucl-th/9910012} {arXiv:nucl-th/9910012} \BibitemShut {NoStop}%
\bibitem [{\citenamefont {Olive}\ \emph {et~al.}(2014)\citenamefont {Olive} \emph {et~al.}}]{ParticleDataGroup:2014cgo}%
  \BibitemOpen
  \bibfield  {author} {\bibinfo {author} {\bibfnamefont {K.~A.}\ \bibnamefont {Olive}} \emph {et~al.} (\bibinfo {collaboration} {Particle Data Group}),\ }\bibfield  {title} {\bibinfo {title} {{Review of Particle Physics}},\ }\href {https://doi.org/10.1088/1674-1137/38/9/090001} {\bibfield  {journal} {\bibinfo  {journal} {Chin. Phys. C}\ }\textbf {\bibinfo {volume} {38}},\ \bibinfo {pages} {090001} (\bibinfo {year} {2014})}\BibitemShut {NoStop}%
\bibitem [{\citenamefont {Bloom}\ and\ \citenamefont {Peck}(1983)}]{Bloom:1983pc}%
  \BibitemOpen
  \bibfield  {author} {\bibinfo {author} {\bibfnamefont {E.~D.}\ \bibnamefont {Bloom}}\ and\ \bibinfo {author} {\bibfnamefont {C.}~\bibnamefont {Peck}},\ }\bibfield  {title} {\bibinfo {title} {{Physics with the Crystal Ball Detector}},\ }\href {https://doi.org/10.1146/annurev.ns.33.120183.001043} {\bibfield  {journal} {\bibinfo  {journal} {Ann. Rev. Nucl. Part. Sci.}\ }\textbf {\bibinfo {volume} {33}},\ \bibinfo {pages} {143} (\bibinfo {year} {1983})}\BibitemShut {NoStop}%
\bibitem [{\citenamefont {Bienlein}(1991)}]{Bienlein:1991jb}%
  \BibitemOpen
  \bibfield  {author} {\bibinfo {author} {\bibfnamefont {J.~K.}\ \bibnamefont {Bienlein}},\ }\href@noop {} {\bibinfo {title} {{The Crystal Ball detector at DORIS-II: Review of achievements}}} (\bibinfo {year} {1991}),\ \bibinfo {note} {internal Report, DESY-F31-91-02}\BibitemShut {NoStop}%
\bibitem [{\citenamefont {Nefkens}(2005)}]{Nefkens:2005an}%
  \BibitemOpen
  \bibfield  {author} {\bibinfo {author} {\bibfnamefont {B.~M.~K.}\ \bibnamefont {Nefkens}},\ }\bibfield  {title} {\bibinfo {title} {{Crystal Ball physics}},\ }\href {https://doi.org/10.1016/j.ppnp.2005.01.020} {\bibfield  {journal} {\bibinfo  {journal} {Prog. Part. Nucl. Phys.}\ ,\ \bibinfo {pages} {153}} (\bibinfo {year} {2005})}\BibitemShut {NoStop}%
\bibitem [{\citenamefont {Denig}(2016)}]{Denig:2016dqo}%
  \BibitemOpen
  \bibfield  {author} {\bibinfo {author} {\bibfnamefont {A.}~\bibnamefont {Denig}},\ }\bibfield  {title} {\bibinfo {title} {{Recent results from the Mainz Microtron MAMI and an outlook for the future}},\ }\href {https://doi.org/10.1063/1.4949374} {\bibfield  {journal} {\bibinfo  {journal} {AIP Conf. Proc.}\ }\textbf {\bibinfo {volume} {1735}},\ \bibinfo {pages} {020006} (\bibinfo {year} {2016})}\BibitemShut {NoStop}%
\bibitem [{\citenamefont {Prakhov}\ \emph {et~al.}(2004)\citenamefont {Prakhov} \emph {et~al.}}]{CrystalBall:2004qln}%
  \BibitemOpen
  \bibfield  {author} {\bibinfo {author} {\bibfnamefont {S.}~\bibnamefont {Prakhov}} \emph {et~al.} (\bibinfo {collaboration} {Crystal Ball}),\ }\bibfield  {title} {\bibinfo {title} {{Measurement of $\pi^- p \to \pi^0 \pi^0 n$ from threshold to $p_{\pi^-}= 750$\,MeV/c}},\ }\href {https://doi.org/10.1103/PhysRevC.69.045202} {\bibfield  {journal} {\bibinfo  {journal} {Phys. Rev. C}\ }\textbf {\bibinfo {volume} {69}},\ \bibinfo {pages} {045202} (\bibinfo {year} {2004})}\BibitemShut {NoStop}%
\bibitem [{\citenamefont {Kamano}\ and\ \citenamefont {Arima}(2006)}]{Kamano:2006vm}%
  \BibitemOpen
  \bibfield  {author} {\bibinfo {author} {\bibfnamefont {H.}~\bibnamefont {Kamano}}\ and\ \bibinfo {author} {\bibfnamefont {M.}~\bibnamefont {Arima}},\ }\bibfield  {title} {\bibinfo {title} {{The $\pi N \to \pi \pi N$ reaction around $N^*(1440)$ energy}},\ }\href {https://doi.org/10.1103/PhysRevC.73.055203} {\bibfield  {journal} {\bibinfo  {journal} {Phys. Rev. C}\ }\textbf {\bibinfo {volume} {73}},\ \bibinfo {pages} {055203} (\bibinfo {year} {2006})},\ \Eprint {https://arxiv.org/abs/nucl-th/0601057} {arXiv:nucl-th/0601057} \BibitemShut {NoStop}%
\bibitem [{\citenamefont {Schneider}\ \emph {et~al.}(2006)\citenamefont {Schneider}, \citenamefont {Krewald},\ and\ \citenamefont {Meissner}}]{Schneider:2006bd}%
  \BibitemOpen
  \bibfield  {author} {\bibinfo {author} {\bibfnamefont {S.}~\bibnamefont {Schneider}}, \bibinfo {author} {\bibfnamefont {S.}~\bibnamefont {Krewald}},\ and\ \bibinfo {author} {\bibfnamefont {U.-G.}\ \bibnamefont {Meissner}},\ }\bibfield  {title} {\bibinfo {title} {{The Reaction $\pi N \to \pi \pi N$ in a meson-exchange approach}},\ }\href {https://doi.org/10.1140/epja/i2006-10030-0} {\bibfield  {journal} {\bibinfo  {journal} {Eur. Phys. J. A}\ }\textbf {\bibinfo {volume} {28}},\ \bibinfo {pages} {107} (\bibinfo {year} {2006})},\ \Eprint {https://arxiv.org/abs/nucl-th/0603040} {arXiv:nucl-th/0603040} \BibitemShut {NoStop}%
\bibitem [{\citenamefont {Siemens}\ \emph {et~al.}(2014)\citenamefont {Siemens} \emph {et~al.}}]{Siemens:2014pma}%
  \BibitemOpen
  \bibfield  {author} {\bibinfo {author} {\bibfnamefont {D.}~\bibnamefont {Siemens}} \emph {et~al.},\ }\bibfield  {title} {\bibinfo {title} {{The reaction $\pi N \to \pi \pi N$ in chiral effective field theory with explicit $\Delta$(1232) degrees of freedom}},\ }\href {https://doi.org/10.1103/PhysRevC.89.065211} {\bibfield  {journal} {\bibinfo  {journal} {Phys. Rev. C}\ }\textbf {\bibinfo {volume} {89}},\ \bibinfo {pages} {065211} (\bibinfo {year} {2014})},\ \Eprint {https://arxiv.org/abs/1403.2510} {arXiv:1403.2510 [nucl-th]} \BibitemShut {NoStop}%
\bibitem [{\citenamefont {Shklyar}\ \emph {et~al.}(2016)\citenamefont {Shklyar}, \citenamefont {Lenske},\ and\ \citenamefont {Mosel}}]{Shklyar:2014kra}%
  \BibitemOpen
  \bibfield  {author} {\bibinfo {author} {\bibfnamefont {V.}~\bibnamefont {Shklyar}}, \bibinfo {author} {\bibfnamefont {H.}~\bibnamefont {Lenske}},\ and\ \bibinfo {author} {\bibfnamefont {U.}~\bibnamefont {Mosel}},\ }\bibfield  {title} {\bibinfo {title} {{$2\pi$ production in the Giessen coupled-channel model}},\ }\href {https://doi.org/10.1103/PhysRevC.93.045206} {\bibfield  {journal} {\bibinfo  {journal} {Phys. Rev. C}\ }\textbf {\bibinfo {volume} {93}},\ \bibinfo {pages} {045206} (\bibinfo {year} {2016})},\ \Eprint {https://arxiv.org/abs/1409.7920} {arXiv:1409.7920 [nucl-th]} \BibitemShut {NoStop}%
\bibitem [{\citenamefont {Kamano}\ \emph {et~al.}(2009)\citenamefont {Kamano} \emph {et~al.}}]{Kamano:2008gr}%
  \BibitemOpen
  \bibfield  {author} {\bibinfo {author} {\bibfnamefont {H.}~\bibnamefont {Kamano}} \emph {et~al.},\ }\bibfield  {title} {\bibinfo {title} {{Dynamical coupled-channels study of $\pi N \to \pi \pi N$ reactions}},\ }\href {https://doi.org/10.1103/PhysRevC.79.025206} {\bibfield  {journal} {\bibinfo  {journal} {Phys. Rev. C}\ }\textbf {\bibinfo {volume} {79}},\ \bibinfo {pages} {025206} (\bibinfo {year} {2009})},\ \Eprint {https://arxiv.org/abs/0807.2273} {arXiv:0807.2273 [nucl-th]} \BibitemShut {NoStop}%
\bibitem [{\citenamefont {Kamano}(2013)}]{Kamano:2013ona}%
  \BibitemOpen
  \bibfield  {author} {\bibinfo {author} {\bibfnamefont {H.}~\bibnamefont {Kamano}},\ }\bibfield  {title} {\bibinfo {title} {{Impact of $\pi N\to \pi \pi N$ data on determining high-mass nucleon resonances}},\ }\href {https://doi.org/10.1103/PhysRevC.88.045203} {\bibfield  {journal} {\bibinfo  {journal} {Phys. Rev. C}\ }\textbf {\bibinfo {volume} {88}},\ \bibinfo {pages} {045203} (\bibinfo {year} {2013})},\ \Eprint {https://arxiv.org/abs/1305.6678} {arXiv:1305.6678 [nucl-th]} \BibitemShut {NoStop}%
\bibitem [{\citenamefont {Adamczewski-Musch}\ \emph {et~al.}(2020)\citenamefont {Adamczewski-Musch} \emph {et~al.}}]{HADES:2020kce}%
  \BibitemOpen
  \bibfield  {author} {\bibinfo {author} {\bibfnamefont {J.}~\bibnamefont {Adamczewski-Musch}} \emph {et~al.} (\bibinfo {collaboration} {HADES}),\ }\bibfield  {title} {\bibinfo {title} {{Two-pion production in the second resonance region in ${\pi}^-p$ collisions with the High-Acceptance Di-Electron Spectrometer (HADES)}},\ }\href {https://doi.org/10.1103/PhysRevC.102.024001} {\bibfield  {journal} {\bibinfo  {journal} {Phys. Rev. C}\ }\textbf {\bibinfo {volume} {102}},\ \bibinfo {pages} {024001} (\bibinfo {year} {2020})},\ \Eprint {https://arxiv.org/abs/2004.08265} {arXiv:2004.08265 [nucl-ex]} \BibitemShut {NoStop}%
\bibitem [{\citenamefont {Crouch}\ \emph {et~al.}(1968)\citenamefont {Crouch} \emph {et~al.}}]{CambridgeBubbleChamberGroup:1968zz}%
  \BibitemOpen
  \bibfield  {author} {\bibinfo {author} {\bibfnamefont {H.~R.}\ \bibnamefont {Crouch}} \emph {et~al.},\ }\bibfield  {title} {\bibinfo {title} {{Multipion Photoproduction at Energies up to 6 BeV}},\ }\href {https://doi.org/10.1103/PhysRev.169.1081} {\bibfield  {journal} {\bibinfo  {journal} {Phys. Rev.}\ }\textbf {\bibinfo {volume} {169}},\ \bibinfo {pages} {1081} (\bibinfo {year} {1968})}\BibitemShut {NoStop}%
\bibitem [{\citenamefont {Erbe}\ \emph {et~al.}(1968{\natexlab{a}})\citenamefont {Erbe} \emph {et~al.}}]{Aachen-Berlin-Bonn-Hamburg-Heidelberg-Munchen:1968dpq}%
  \BibitemOpen
  \bibfield  {author} {\bibinfo {author} {\bibfnamefont {R.}~\bibnamefont {Erbe}} \emph {et~al.} (\bibinfo {collaboration} {Aachen-Berlin-Bonn-Hamburg-Heidelberg-M\"unchen}),\ }\bibfield  {title} {\bibinfo {title} {{Photoproduction of vector mesons on protons at energies up to 5.8 GeV}},\ }\href {https://doi.org/10.1016/0370-2693(68)90332-8} {\bibfield  {journal} {\bibinfo  {journal} {Phys. Lett. B}\ }\textbf {\bibinfo {volume} {27}},\ \bibinfo {pages} {54} (\bibinfo {year} {1968}{\natexlab{a}})}\BibitemShut {NoStop}%
\bibitem [{\citenamefont {Ballam}\ \emph {et~al.}(1972{\natexlab{a}})\citenamefont {Ballam} \emph {et~al.}}]{Ballam:1971wq}%
  \BibitemOpen
  \bibfield  {author} {\bibinfo {author} {\bibfnamefont {J.}~\bibnamefont {Ballam}} \emph {et~al.},\ }\bibfield  {title} {\bibinfo {title} {{Study of high-energy photoproduction with positron annihilation radiation.I. Three prong events}},\ }\href {https://doi.org/10.1103/PhysRevD.5.15} {\bibfield  {journal} {\bibinfo  {journal} {Phys. Rev. D}\ }\textbf {\bibinfo {volume} {5}},\ \bibinfo {pages} {15} (\bibinfo {year} {1972}{\natexlab{a}})}\BibitemShut {NoStop}%
\bibitem [{\citenamefont {Ballam}\ \emph {et~al.}(1972{\natexlab{b}})\citenamefont {Ballam} \emph {et~al.}}]{Ballam:1971yd}%
  \BibitemOpen
  \bibfield  {author} {\bibinfo {author} {\bibfnamefont {J.}~\bibnamefont {Ballam}} \emph {et~al.},\ }\bibfield  {title} {\bibinfo {title} {{Bubble Chamber Study of Photoproduction by 2.8\,GeV and 4.7\,GeV Polarized Photons. 1. Cross-Section Determinations and Production of $\rho^0$ and $\Delta^{++}$ in the Reaction $\gamma p \to p \pi^+ \pi^-$}},\ }\href {https://doi.org/10.1103/PhysRevD.5.545} {\bibfield  {journal} {\bibinfo  {journal} {Phys. Rev. D}\ }\textbf {\bibinfo {volume} {5}},\ \bibinfo {pages} {545} (\bibinfo {year} {1972}{\natexlab{b}})}\BibitemShut {NoStop}%
\bibitem [{\citenamefont {Lanzerotti}(1965)}]{Lanzerotti:1965bp}%
  \BibitemOpen
  \bibfield  {author} {\bibinfo {author} {\bibfnamefont {L.~J.}\ \bibnamefont {Lanzerotti}},\ }\emph {\bibinfo {title} {{Photoproduction of charged pion pairs}}},\ \href@noop {} {\bibinfo {type} {Harvard thesis}} (\bibinfo {year} {1965})\BibitemShut {NoStop}%
\bibitem [{\citenamefont {Lanzerotti}\ \emph {et~al.}(1968)\citenamefont {Lanzerotti} \emph {et~al.}}]{Lanzerotti:1968zz}%
  \BibitemOpen
  \bibfield  {author} {\bibinfo {author} {\bibfnamefont {L.~J.}\ \bibnamefont {Lanzerotti}} \emph {et~al.},\ }\bibfield  {title} {\bibinfo {title} {{Photoproduction of Neutral $\rho$ Mesons}},\ }\href {https://doi.org/10.1103/PhysRev.166.1365} {\bibfield  {journal} {\bibinfo  {journal} {Phys. Rev.}\ }\textbf {\bibinfo {volume} {166}},\ \bibinfo {pages} {1365} (\bibinfo {year} {1968})}\BibitemShut {NoStop}%
\bibitem [{\citenamefont {Söding}(1966)}]{Soding:1965nh}%
  \BibitemOpen
  \bibfield  {author} {\bibinfo {author} {\bibfnamefont {P.}~\bibnamefont {Söding}},\ }\bibfield  {title} {\bibinfo {title} {{On the Apparent shift of the $\rho$ meson mass in photoproduction}},\ }\href {https://doi.org/10.1016/0031-9163(66)90451-3} {\bibfield  {journal} {\bibinfo  {journal} {Phys. Lett.}\ }\textbf {\bibinfo {volume} {19}},\ \bibinfo {pages} {702} (\bibinfo {year} {1966})}\BibitemShut {NoStop}%
\bibitem [{\citenamefont {Kroll}\ and\ \citenamefont {Ruderman}(1954)}]{Kroll:1954}%
  \BibitemOpen
  \bibfield  {author} {\bibinfo {author} {\bibfnamefont {N.~M.}\ \bibnamefont {Kroll}}\ and\ \bibinfo {author} {\bibfnamefont {M.~A.}\ \bibnamefont {Ruderman}},\ }\bibfield  {title} {\bibinfo {title} {A theorem on photomeson production near threshold and the suppression of pairs in pseudoscalar meson theory},\ }\href@noop {} {\bibfield  {journal} {\bibinfo  {journal} {Phys. Rev.}\ }\textbf {\bibinfo {volume} {93}},\ \bibinfo {pages} {233} (\bibinfo {year} {1954})}\BibitemShut {NoStop}%
\bibitem [{\citenamefont {Struczinski}\ \emph {et~al.}(1976)\citenamefont {Struczinski} \emph {et~al.}}]{Aachen-Hamburg-Heidelberg-Munich:1975jed}%
  \BibitemOpen
  \bibfield  {author} {\bibinfo {author} {\bibfnamefont {W.}~\bibnamefont {Struczinski}} \emph {et~al.} (\bibinfo {collaboration} {Aachen-Hamburg-Heidelberg-Munich}),\ }\bibfield  {title} {\bibinfo {title} {{Study of photoproduction on hydrogen in a streamer chamber with tagged photons for 1.6 GeV $ < E_\gamma <$ 6.3 GeV Topological and reaction cross sections}},\ }\href {https://doi.org/10.1016/0550-3213(76)90123-1} {\bibfield  {journal} {\bibinfo  {journal} {Nucl. Phys. B}\ }\textbf {\bibinfo {volume} {108}},\ \bibinfo {pages} {45} (\bibinfo {year} {1976})}\BibitemShut {NoStop}%
\bibitem [{\citenamefont {Aubert}\ \emph {et~al.}(1974)\citenamefont {Aubert} \emph {et~al.}}]{E598:1974sol}%
  \BibitemOpen
  \bibfield  {author} {\bibinfo {author} {\bibfnamefont {J.~J.}\ \bibnamefont {Aubert}} \emph {et~al.} (\bibinfo {collaboration} {E598}),\ }\bibfield  {title} {\bibinfo {title} {{Experimental Observation of a Heavy Particle $J$}},\ }\href {https://doi.org/10.1103/PhysRevLett.33.1404} {\bibfield  {journal} {\bibinfo  {journal} {Phys. Rev. Lett.}\ }\textbf {\bibinfo {volume} {33}},\ \bibinfo {pages} {1404} (\bibinfo {year} {1974})}\BibitemShut {NoStop}%
\bibitem [{\citenamefont {Augustin}\ \emph {et~al.}(1974)\citenamefont {Augustin} \emph {et~al.}}]{SLAC-SP-017:1974ind}%
  \BibitemOpen
  \bibfield  {author} {\bibinfo {author} {\bibfnamefont {J.~E.}\ \bibnamefont {Augustin}} \emph {et~al.} (\bibinfo {collaboration} {SLAC-SP-017}),\ }\bibfield  {title} {\bibinfo {title} {{Discovery of a Narrow Resonance in $e^+ e^-$ Annihilation}},\ }\href {https://doi.org/10.1103/PhysRevLett.33.1406} {\bibfield  {journal} {\bibinfo  {journal} {Phys. Rev. Lett.}\ }\textbf {\bibinfo {volume} {33}},\ \bibinfo {pages} {1406} (\bibinfo {year} {1974})}\BibitemShut {NoStop}%
\bibitem [{\citenamefont {Köpke}\ and\ \citenamefont {Wermes}(1989)}]{Kopke:1988cs}%
  \BibitemOpen
  \bibfield  {author} {\bibinfo {author} {\bibfnamefont {L.}~\bibnamefont {Köpke}}\ and\ \bibinfo {author} {\bibfnamefont {N.}~\bibnamefont {Wermes}},\ }\bibfield  {title} {\bibinfo {title} {{$J/\psi$ Decays}},\ }\href {https://doi.org/10.1016/0370-1573(89)90074-4} {\bibfield  {journal} {\bibinfo  {journal} {Phys. Rept.}\ }\textbf {\bibinfo {volume} {174}},\ \bibinfo {pages} {67} (\bibinfo {year} {1989})}\BibitemShut {NoStop}%
\bibitem [{\citenamefont {Weinberg}(1979)}]{Weinberg:1978kz}%
  \BibitemOpen
  \bibfield  {author} {\bibinfo {author} {\bibfnamefont {S.}~\bibnamefont {Weinberg}},\ }\bibfield  {title} {\bibinfo {title} {{Phenomenological Lagrangians}},\ }\href {https://doi.org/10.1016/0378-4371(79)90223-1} {\bibfield  {journal} {\bibinfo  {journal} {Physica A}\ }\textbf {\bibinfo {volume} {96}},\ \bibinfo {pages} {327} (\bibinfo {year} {1979})}\BibitemShut {NoStop}%
\bibitem [{\citenamefont {Gasser}\ and\ \citenamefont {Leutwyler}(1984)}]{Gasser:1983yg}%
  \BibitemOpen
  \bibfield  {author} {\bibinfo {author} {\bibfnamefont {J.}~\bibnamefont {Gasser}}\ and\ \bibinfo {author} {\bibfnamefont {H.}~\bibnamefont {Leutwyler}},\ }\bibfield  {title} {\bibinfo {title} {{Chiral Perturbation Theory to One Loop}},\ }\href {https://doi.org/10.1016/0003-4916(84)90242-2} {\bibfield  {journal} {\bibinfo  {journal} {Annals Phys.}\ }\textbf {\bibinfo {volume} {158}},\ \bibinfo {pages} {142} (\bibinfo {year} {1984})}\BibitemShut {NoStop}%
\bibitem [{\citenamefont {Braghieri}\ \emph {et~al.}(1995)\citenamefont {Braghieri} \emph {et~al.}}]{Braghieri:1994rf}%
  \BibitemOpen
  \bibfield  {author} {\bibinfo {author} {\bibfnamefont {A.}~\bibnamefont {Braghieri}} \emph {et~al.},\ }\bibfield  {title} {\bibinfo {title} {{Total cross-section measurement for the three double pion production channels on the proton}},\ }\href {https://doi.org/10.1016/0370-2693(95)01189-W} {\bibfield  {journal} {\bibinfo  {journal} {Phys. Lett. B}\ }\textbf {\bibinfo {volume} {363}},\ \bibinfo {pages} {46} (\bibinfo {year} {1995})}\BibitemShut {NoStop}%
\bibitem [{\citenamefont {Harter}\ \emph {et~al.}(1997)\citenamefont {Harter} \emph {et~al.}}]{Harter:1997jq}%
  \BibitemOpen
  \bibfield  {author} {\bibinfo {author} {\bibfnamefont {F.}~\bibnamefont {Harter}} \emph {et~al.},\ }\bibfield  {title} {\bibinfo {title} {{Two neutral pion photoproduction off the proton between threshold and 800\,MeV}},\ }\href {https://doi.org/10.1016/S0370-2693(97)00423-1} {\bibfield  {journal} {\bibinfo  {journal} {Phys. Lett. B}\ }\textbf {\bibinfo {volume} {401}},\ \bibinfo {pages} {229} (\bibinfo {year} {1997})}\BibitemShut {NoStop}%
\bibitem [{\citenamefont {Wolf}\ \emph {et~al.}(2000)\citenamefont {Wolf} \emph {et~al.}}]{Wolf:2000qt}%
  \BibitemOpen
  \bibfield  {author} {\bibinfo {author} {\bibfnamefont {M.}~\bibnamefont {Wolf}} \emph {et~al.},\ }\bibfield  {title} {\bibinfo {title} {{Photoproduction of neutral pion pairs from the proton}},\ }\href {https://doi.org/10.1007/s100500070048} {\bibfield  {journal} {\bibinfo  {journal} {Eur. Phys. J. A}\ }\textbf {\bibinfo {volume} {9}},\ \bibinfo {pages} {5} (\bibinfo {year} {2000})}\BibitemShut {NoStop}%
\bibitem [{\citenamefont {Langgartner}\ \emph {et~al.}(2001)\citenamefont {Langgartner} \emph {et~al.}}]{Langgartner:2001sg}%
  \BibitemOpen
  \bibfield  {author} {\bibinfo {author} {\bibfnamefont {W.}~\bibnamefont {Langgartner}} \emph {et~al.},\ }\bibfield  {title} {\bibinfo {title} {{Direct observation of a $\rho$ decay of the $D_{13}(1520)$ baryon resonance}},\ }\href {https://doi.org/10.1103/PhysRevLett.87.052001} {\bibfield  {journal} {\bibinfo  {journal} {Phys. Rev. Lett.}\ }\textbf {\bibinfo {volume} {87}},\ \bibinfo {pages} {052001} (\bibinfo {year} {2001})}\BibitemShut {NoStop}%
\bibitem [{\citenamefont {Kotulla}\ \emph {et~al.}(2004)\citenamefont {Kotulla} \emph {et~al.}}]{Kotulla:2003cx}%
  \BibitemOpen
  \bibfield  {author} {\bibinfo {author} {\bibfnamefont {M.}~\bibnamefont {Kotulla}} \emph {et~al.},\ }\bibfield  {title} {\bibinfo {title} {{Double $\pi^0$ photoproduction off the proton at threshold}},\ }\href {https://doi.org/10.1016/j.physletb.2003.10.056} {\bibfield  {journal} {\bibinfo  {journal} {Phys. Lett. B}\ }\textbf {\bibinfo {volume} {578}},\ \bibinfo {pages} {63} (\bibinfo {year} {2004})},\ \Eprint {https://arxiv.org/abs/nucl-ex/0310031} {arXiv:nucl-ex/0310031} \BibitemShut {NoStop}%
\bibitem [{\citenamefont {Ahrens}\ \emph {et~al.}(2005)\citenamefont {Ahrens} \emph {et~al.}}]{GDH:2005jgl}%
  \BibitemOpen
  \bibfield  {author} {\bibinfo {author} {\bibfnamefont {J.}~\bibnamefont {Ahrens}} \emph {et~al.} (\bibinfo {collaboration} {GDH, A2}),\ }\bibfield  {title} {\bibinfo {title} {{Intermediate resonance excitation in the $\gamma p \to p \pi^0 \pi^0$ reaction}},\ }\href {https://doi.org/10.1016/j.physletb.2005.08.034} {\bibfield  {journal} {\bibinfo  {journal} {Phys. Lett. B}\ }\textbf {\bibinfo {volume} {624}},\ \bibinfo {pages} {173} (\bibinfo {year} {2005})}\BibitemShut {NoStop}%
\bibitem [{\citenamefont {Ahrens}\ \emph {et~al.}(2007)\citenamefont {Ahrens} \emph {et~al.}}]{GDH:2007nkn}%
  \BibitemOpen
  \bibfield  {author} {\bibinfo {author} {\bibfnamefont {J.}~\bibnamefont {Ahrens}} \emph {et~al.} (\bibinfo {collaboration} {GDH, A2}),\ }\bibfield  {title} {\bibinfo {title} {{First measurement of the helicity dependence for the $\gamma p \to p \pi^+ \pi^-$ reaction}},\ }\href {https://doi.org/10.1140/epja/i2007-10491-5} {\bibfield  {journal} {\bibinfo  {journal} {Eur. Phys. J. A}\ }\textbf {\bibinfo {volume} {34}},\ \bibinfo {pages} {11} (\bibinfo {year} {2007})}\BibitemShut {NoStop}%
\bibitem [{\citenamefont {Krambrich}\ \emph {et~al.}(2009)\citenamefont {Krambrich} \emph {et~al.}}]{CrystalBallatMAMI:2009iym}%
  \BibitemOpen
  \bibfield  {author} {\bibinfo {author} {\bibfnamefont {D.}~\bibnamefont {Krambrich}} \emph {et~al.} (\bibinfo {collaboration} {Crystal Ball at MAMI, TAPS, A2}),\ }\bibfield  {title} {\bibinfo {title} {{Beam-Helicity Asymmetries in Double Pion Photoproduction off the Proton}},\ }\href {https://doi.org/10.1103/PhysRevLett.103.052002} {\bibfield  {journal} {\bibinfo  {journal} {Phys. Rev. Lett.}\ }\textbf {\bibinfo {volume} {103}},\ \bibinfo {pages} {052002} (\bibinfo {year} {2009})},\ \Eprint {https://arxiv.org/abs/0907.0358} {arXiv:0907.0358 [nucl-ex]} \BibitemShut {NoStop}%
\bibitem [{\citenamefont {Gomez~Tejedor}\ and\ \citenamefont {Oset}(1996)}]{GomezTejedor:1995pe}%
  \BibitemOpen
  \bibfield  {author} {\bibinfo {author} {\bibfnamefont {J.~A.}\ \bibnamefont {Gomez~Tejedor}}\ and\ \bibinfo {author} {\bibfnamefont {E.}~\bibnamefont {Oset}},\ }\bibfield  {title} {\bibinfo {title} {{Double pion photoproduction on the nucleon: Study of the isospin channels}},\ }\href {https://doi.org/10.1016/0375-9474(95)00492-0} {\bibfield  {journal} {\bibinfo  {journal} {Nucl. Phys. A}\ }\textbf {\bibinfo {volume} {600}},\ \bibinfo {pages} {413} (\bibinfo {year} {1996})},\ \Eprint {https://arxiv.org/abs/hep-ph/9506209} {arXiv:hep-ph/9506209} \BibitemShut {NoStop}%
\bibitem [{\citenamefont {Gomez~Tejedor}\ \emph {et~al.}(1996)\citenamefont {Gomez~Tejedor}, \citenamefont {Cano},\ and\ \citenamefont {Oset}}]{GomezTejedor:1995kj}%
  \BibitemOpen
  \bibfield  {author} {\bibinfo {author} {\bibfnamefont {J.~A.}\ \bibnamefont {Gomez~Tejedor}}, \bibinfo {author} {\bibfnamefont {F.}~\bibnamefont {Cano}},\ and\ \bibinfo {author} {\bibfnamefont {E.}~\bibnamefont {Oset}},\ }\bibfield  {title} {\bibinfo {title} {{The $N^*(1520) \to\Delta\pi$ amplitudes extracted from the $\gamma p\to\pi^+\pi^- p$ reaction and comparison to quark models}},\ }\href {https://doi.org/10.1016/0370-2693(96)00404-2} {\bibfield  {journal} {\bibinfo  {journal} {Phys. Lett. B}\ }\textbf {\bibinfo {volume} {379}},\ \bibinfo {pages} {39} (\bibinfo {year} {1996})},\ \Eprint {https://arxiv.org/abs/nucl-th/9510007} {arXiv:nucl-th/9510007} \BibitemShut {NoStop}%
\bibitem [{\citenamefont {Kashevarov}\ \emph {et~al.}(2012)\citenamefont {Kashevarov} \emph {et~al.}}]{Kashevarov:2012wy}%
  \BibitemOpen
  \bibfield  {author} {\bibinfo {author} {\bibfnamefont {V.~L.}\ \bibnamefont {Kashevarov}} \emph {et~al.},\ }\bibfield  {title} {\bibinfo {title} {{Study of the $\gamma p \to \pi^0 \pi^0 p$ reaction with the Crystal Ball/TAPS at the Mainz}},\ }\href {https://doi.org/10.1103/PhysRevC.85.064610} {\bibfield  {journal} {\bibinfo  {journal} {Phys. Rev. C}\ }\textbf {\bibinfo {volume} {85}},\ \bibinfo {pages} {064610} (\bibinfo {year} {2012})},\ \Eprint {https://arxiv.org/abs/1204.4058} {arXiv:1204.4058 [nucl-ex]} \BibitemShut {NoStop}%
\bibitem [{\citenamefont {Zehr}\ \emph {et~al.}(2012)\citenamefont {Zehr} \emph {et~al.}}]{Zehr:2012tj}%
  \BibitemOpen
  \bibfield  {author} {\bibinfo {author} {\bibfnamefont {F.}~\bibnamefont {Zehr}} \emph {et~al.},\ }\bibfield  {title} {\bibinfo {title} {{Photoproduction of $\pi^0\pi^0$ and $\pi^0\pi^\pm$ pairs off the proton from threshold to the second resonance region}},\ }\href {https://doi.org/10.1140/epja/i2012-12098-1} {\bibfield  {journal} {\bibinfo  {journal} {Eur. Phys. J. A}\ }\textbf {\bibinfo {volume} {48}},\ \bibinfo {pages} {98} (\bibinfo {year} {2012})},\ \Eprint {https://arxiv.org/abs/1207.2361} {arXiv:1207.2361 [nucl-ex]} \BibitemShut {NoStop}%
\bibitem [{\citenamefont {Oberle}\ \emph {et~al.}(2013)\citenamefont {Oberle} \emph {et~al.}}]{Oberle:2013kvb}%
  \BibitemOpen
  \bibfield  {author} {\bibinfo {author} {\bibfnamefont {M.}~\bibnamefont {Oberle}} \emph {et~al.},\ }\bibfield  {title} {\bibinfo {title} {{Measurement of the beam-helicity asymmetry $I^{\odot}$ in the photoproduction of $\pi^0$-pairs off the proton and off the neutron}},\ }\href {https://doi.org/10.1016/j.physletb.2013.03.021} {\bibfield  {journal} {\bibinfo  {journal} {Phys. Lett. B}\ }\textbf {\bibinfo {volume} {721}},\ \bibinfo {pages} {237} (\bibinfo {year} {2013})},\ \Eprint {https://arxiv.org/abs/1304.1919} {arXiv:1304.1919 [nucl-ex]} \BibitemShut {NoStop}%
\bibitem [{\citenamefont {Oberle}\ \emph {et~al.}(2014)\citenamefont {Oberle} \emph {et~al.}}]{A2:2014snn}%
  \BibitemOpen
  \bibfield  {author} {\bibinfo {author} {\bibfnamefont {M.}~\bibnamefont {Oberle}} \emph {et~al.} (\bibinfo {collaboration} {A2, MAMI, TAPS}),\ }\bibfield  {title} {\bibinfo {title} {{Measurement of the beam-helicity asymmetry $I^\odot$ in the photoproduction of $\pi^0\pi^{\pm}$ pairs off protons and off neutrons}},\ }\href {https://doi.org/10.1140/epja/i2014-14054-5} {\bibfield  {journal} {\bibinfo  {journal} {Eur. Phys. J. A}\ }\textbf {\bibinfo {volume} {50}},\ \bibinfo {pages} {54} (\bibinfo {year} {2014})},\ \Eprint {https://arxiv.org/abs/1403.1989} {arXiv:1403.1989 [nucl-ex]} \BibitemShut {NoStop}%
\bibitem [{\citenamefont {Dieterle}\ \emph {et~al.}(2015)\citenamefont {Dieterle} \emph {et~al.}}]{A2:2015pgk}%
  \BibitemOpen
  \bibfield  {author} {\bibinfo {author} {\bibfnamefont {M.}~\bibnamefont {Dieterle}} \emph {et~al.} (\bibinfo {collaboration} {A2}),\ }\bibfield  {title} {\bibinfo {title} {{Photoproduction of $ \pi^{0}$ -pairs off protons and off neutrons}},\ }\href {https://doi.org/10.1140/epja/i2015-15142-8} {\bibfield  {journal} {\bibinfo  {journal} {Eur. Phys. J. A}\ }\textbf {\bibinfo {volume} {51}},\ \bibinfo {pages} {142} (\bibinfo {year} {2015})},\ \Eprint {https://arxiv.org/abs/1510.09167} {arXiv:1510.09167 [nucl-ex]} \BibitemShut {NoStop}%
\bibitem [{\citenamefont {Assafiri}\ \emph {et~al.}(2003)\citenamefont {Assafiri} \emph {et~al.}}]{Assafiri:2003mv}%
  \BibitemOpen
  \bibfield  {author} {\bibinfo {author} {\bibfnamefont {Y.}~\bibnamefont {Assafiri}} \emph {et~al.},\ }\bibfield  {title} {\bibinfo {title} {{Double $\pi^0$ photoproduction on the proton at GRAAL}},\ }\href {https://doi.org/10.1103/PhysRevLett.90.222001} {\bibfield  {journal} {\bibinfo  {journal} {Phys. Rev. Lett.}\ }\textbf {\bibinfo {volume} {90}},\ \bibinfo {pages} {222001} (\bibinfo {year} {2003})}\BibitemShut {NoStop}%
\bibitem [{\citenamefont {Wu}\ \emph {et~al.}(2005)\citenamefont {Wu} \emph {et~al.}}]{Wu:2005wf}%
  \BibitemOpen
  \bibfield  {author} {\bibinfo {author} {\bibfnamefont {C.}~\bibnamefont {Wu}} \emph {et~al.},\ }\bibfield  {title} {\bibinfo {title} {{Photoproduction of $\rho^0$ mesons and $\Delta$-baryons in the reaction $\gamma p \to p \pi^+ \pi^-$ at energies up to $\sqrt s = 2.6$\,GeV}},\ }\href {https://doi.org/10.1140/epja/i2004-10093-9} {\bibfield  {journal} {\bibinfo  {journal} {Eur. Phys. J. A}\ }\textbf {\bibinfo {volume} {23}},\ \bibinfo {pages} {317} (\bibinfo {year} {2005})}\BibitemShut {NoStop}%
\bibitem [{\citenamefont {Aker}\ \emph {et~al.}(1992)\citenamefont {Aker} \emph {et~al.}}]{CrystalBarrel:1992qav}%
  \BibitemOpen
  \bibfield  {author} {\bibinfo {author} {\bibfnamefont {E.}~\bibnamefont {Aker}} \emph {et~al.} (\bibinfo {collaboration} {Crystal Barrel}),\ }\bibfield  {title} {\bibinfo {title} {{The Crystal Barrel spectrometer at LEAR}},\ }\href {https://doi.org/10.1016/0168-9002(92)90379-I} {\bibfield  {journal} {\bibinfo  {journal} {Nucl. Instrum. Meth. A}\ }\textbf {\bibinfo {volume} {321}},\ \bibinfo {pages} {69} (\bibinfo {year} {1992})}\BibitemShut {NoStop}%
\bibitem [{\citenamefont {Klempt}\ \emph {et~al.}(2002)\citenamefont {Klempt}, \citenamefont {Bradamante}, \citenamefont {Martin},\ and\ \citenamefont {Richard}}]{Klempt:2002ap}%
  \BibitemOpen
  \bibfield  {author} {\bibinfo {author} {\bibfnamefont {E.}~\bibnamefont {Klempt}}, \bibinfo {author} {\bibfnamefont {F.}~\bibnamefont {Bradamante}}, \bibinfo {author} {\bibfnamefont {A.}~\bibnamefont {Martin}},\ and\ \bibinfo {author} {\bibfnamefont {J.~M.}\ \bibnamefont {Richard}},\ }\bibfield  {title} {\bibinfo {title} {{Antinucleon nucleon interaction at low energy: Scattering and protonium}},\ }\href {https://doi.org/10.1016/S0370-1573(02)00144-8} {\bibfield  {journal} {\bibinfo  {journal} {Phys. Rept.}\ }\textbf {\bibinfo {volume} {368}},\ \bibinfo {pages} {119} (\bibinfo {year} {2002})}\BibitemShut {NoStop}%
\bibitem [{\citenamefont {Klempt}\ \emph {et~al.}(2005)\citenamefont {Klempt}, \citenamefont {Batty},\ and\ \citenamefont {Richard}}]{Klempt:2005pp}%
  \BibitemOpen
  \bibfield  {author} {\bibinfo {author} {\bibfnamefont {E.}~\bibnamefont {Klempt}}, \bibinfo {author} {\bibfnamefont {C.}~\bibnamefont {Batty}},\ and\ \bibinfo {author} {\bibfnamefont {J.-M.}\ \bibnamefont {Richard}},\ }\bibfield  {title} {\bibinfo {title} {{The Antinucleon-nucleon interaction at low energy : Annihilation dynamics}},\ }\href {https://doi.org/10.1016/j.physrep.2005.03.002} {\bibfield  {journal} {\bibinfo  {journal} {Phys. Rept.}\ }\textbf {\bibinfo {volume} {413}},\ \bibinfo {pages} {197} (\bibinfo {year} {2005})},\ \Eprint {https://arxiv.org/abs/hep-ex/0501020} {arXiv:hep-ex/0501020} \BibitemShut {NoStop}%
\bibitem [{\citenamefont {Bugg}(2004)}]{Bugg:2004xu}%
  \BibitemOpen
  \bibfield  {author} {\bibinfo {author} {\bibfnamefont {D.~V.}\ \bibnamefont {Bugg}},\ }\bibfield  {title} {\bibinfo {title} {{Four sorts of meson}},\ }\href {https://doi.org/10.1016/j.physrep.2004.03.008} {\bibfield  {journal} {\bibinfo  {journal} {Phys. Rept.}\ }\textbf {\bibinfo {volume} {397}},\ \bibinfo {pages} {257} (\bibinfo {year} {2004})},\ \Eprint {https://arxiv.org/abs/hep-ex/0412045} {arXiv:hep-ex/0412045} \BibitemShut {NoStop}%
\bibitem [{\citenamefont {Amsler}()}]{Amsler:2019ytk}%
  \BibitemOpen
  \bibfield  {author} {\bibinfo {author} {\bibfnamefont {C.}~\bibnamefont {Amsler}},\ }\bibfield  {title} {\bibinfo {title} {{Nucleon-antinucleon annihilation at LEAR}},\ }in\ \href@noop {} {\emph {\bibinfo {booktitle} {ECT* workshop on``Antiproton-nucleus interactions"}}},\ \bibinfo {note} {{Trento 17–21, June 2019}},\ \Eprint {https://arxiv.org/abs/1908.08455} {arXiv:1908.08455 [hep-ph]} \BibitemShut {NoStop}%
\bibitem [{\citenamefont {Arndt}\ \emph {et~al.}(2006)\citenamefont {Arndt}, \citenamefont {Briscoe}, \citenamefont {Strakovsky},\ and\ \citenamefont {Workman}}]{Arndt:2006bf}%
  \BibitemOpen
  \bibfield  {author} {\bibinfo {author} {\bibfnamefont {R.~A.}\ \bibnamefont {Arndt}}, \bibinfo {author} {\bibfnamefont {W.~J.}\ \bibnamefont {Briscoe}}, \bibinfo {author} {\bibfnamefont {I.~I.}\ \bibnamefont {Strakovsky}},\ and\ \bibinfo {author} {\bibfnamefont {R.~L.}\ \bibnamefont {Workman}},\ }\bibfield  {title} {\bibinfo {title} {{Extended partial-wave analysis of $\pi N$ scattering data}},\ }\href {https://doi.org/10.1103/PhysRevC.74.045205} {\bibfield  {journal} {\bibinfo  {journal} {Phys. Rev.}\ }\textbf {\bibinfo {volume} {C74}},\ \bibinfo {pages} {045205} (\bibinfo {year} {2006})},\ \Eprint {https://arxiv.org/abs/nucl-th/0605082} {arXiv:nucl-th/0605082} \BibitemShut {NoStop}%
\bibitem [{\citenamefont {Sarantsev}\ \emph {et~al.}(2008)\citenamefont {Sarantsev} \emph {et~al.}}]{Sarantsev:2007aa}%
  \BibitemOpen
  \bibfield  {author} {\bibinfo {author} {\bibfnamefont {A.~V.}\ \bibnamefont {Sarantsev}} \emph {et~al.},\ }\bibfield  {title} {\bibinfo {title} {{New results on the Roper resonance and the $P_{11}$ partial wave}},\ }\href {https://doi.org/10.1016/j.physletb.2007.11.055} {\bibfield  {journal} {\bibinfo  {journal} {Phys. Lett. B}\ }\textbf {\bibinfo {volume} {659}},\ \bibinfo {pages} {94} (\bibinfo {year} {2008})},\ \Eprint {https://arxiv.org/abs/0707.3591} {arXiv:0707.3591 [hep-ph]} \BibitemShut {NoStop}%
\bibitem [{\citenamefont {Thoma}\ \emph {et~al.}(2008)\citenamefont {Thoma} \emph {et~al.}}]{Thoma:2007bm}%
  \BibitemOpen
  \bibfield  {author} {\bibinfo {author} {\bibfnamefont {U.}~\bibnamefont {Thoma}} \emph {et~al.},\ }\bibfield  {title} {\bibinfo {title} {{$N^*$ and $\Delta^*$ decays into $N \pi^0 \pi^0$}},\ }\href {https://doi.org/10.1016/j.physletb.2007.11.054} {\bibfield  {journal} {\bibinfo  {journal} {Phys. Lett. B}\ }\textbf {\bibinfo {volume} {659}},\ \bibinfo {pages} {87} (\bibinfo {year} {2008})},\ \Eprint {https://arxiv.org/abs/0707.3592} {arXiv:0707.3592 [hep-ph]} \BibitemShut {NoStop}%
\bibitem [{\citenamefont {Sokhoyan}\ \emph {et~al.}(2015{\natexlab{b}})\citenamefont {Sokhoyan} \emph {et~al.}}]{CBELSATAPS:2015tyg}%
  \BibitemOpen
  \bibfield  {author} {\bibinfo {author} {\bibfnamefont {V.}~\bibnamefont {Sokhoyan}} \emph {et~al.} (\bibinfo {collaboration} {CBELSA/TAPS}),\ }\bibfield  {title} {\bibinfo {title} {{Data on $I^s$ and $I^c$ in $\overrightarrow{\gamma}p\to p\pi^0\pi^0$ reveal cascade decays of $N(1900)$ via $N(1520)\pi$}},\ }\href {https://doi.org/10.1016/j.physletb.2015.04.063} {\bibfield  {journal} {\bibinfo  {journal} {Phys. Lett. B}\ }\textbf {\bibinfo {volume} {746}},\ \bibinfo {pages} {127} (\bibinfo {year} {2015}{\natexlab{b}})}\BibitemShut {NoStop}%
\bibitem [{\citenamefont {Thiel}\ \emph {et~al.}(2015)\citenamefont {Thiel} \emph {et~al.}}]{CBELSATAPS:2015taz}%
  \BibitemOpen
  \bibfield  {author} {\bibinfo {author} {\bibfnamefont {A.}~\bibnamefont {Thiel}} \emph {et~al.} (\bibinfo {collaboration} {CBELSA/TAPS}),\ }\bibfield  {title} {\bibinfo {title} {{Three-body nature of $N^{\bf *}$ and $\Delta^*$ resonances from sequential decay chains}},\ }\href {https://doi.org/10.1103/PhysRevLett.114.091803} {\bibfield  {journal} {\bibinfo  {journal} {Phys. Rev. Lett.}\ }\textbf {\bibinfo {volume} {114}},\ \bibinfo {pages} {091803} (\bibinfo {year} {2015})},\ \Eprint {https://arxiv.org/abs/1501.02094} {arXiv:1501.02094 [nucl-ex]} \BibitemShut {NoStop}%
\bibitem [{\citenamefont {Strauch}\ \emph {et~al.}(2005)\citenamefont {Strauch} \emph {et~al.}}]{CLAS:2005oqk}%
  \BibitemOpen
  \bibfield  {author} {\bibinfo {author} {\bibfnamefont {S.}~\bibnamefont {Strauch}} \emph {et~al.} (\bibinfo {collaboration} {CLAS}),\ }\bibfield  {title} {\bibinfo {title} {{Beam-helicity asymmetries in double-charged-pion photoproduction on the proton}},\ }\href {https://doi.org/10.1103/PhysRevLett.95.162003} {\bibfield  {journal} {\bibinfo  {journal} {Phys. Rev. Lett.}\ }\textbf {\bibinfo {volume} {95}},\ \bibinfo {pages} {162003} (\bibinfo {year} {2005})},\ \Eprint {https://arxiv.org/abs/hep-ex/0508002} {arXiv:hep-ex/0508002} \BibitemShut {NoStop}%
\bibitem [{\citenamefont {Koch}(1985)}]{Koch:1985bp}%
  \BibitemOpen
  \bibfield  {author} {\bibinfo {author} {\bibfnamefont {R.}~\bibnamefont {Koch}},\ }\bibfield  {title} {\bibinfo {title} {{Improved $\pi N$ Partial Waves Consistent With Analyticity and Unitarity}},\ }\href {https://doi.org/10.1007/BF01560295} {\bibfield  {journal} {\bibinfo  {journal} {Z. Phys. C}\ }\textbf {\bibinfo {volume} {29}},\ \bibinfo {pages} {597} (\bibinfo {year} {1985})}\BibitemShut {NoStop}%
\bibitem [{\citenamefont {Workman}\ \emph {et~al.}(2012)\citenamefont {Workman}, \citenamefont {Arndt}, \citenamefont {Briscoe}, \citenamefont {Paris},\ and\ \citenamefont {Strakovsky}}]{Workman:2012hx}%
  \BibitemOpen
  \bibfield  {author} {\bibinfo {author} {\bibfnamefont {R.~L.}\ \bibnamefont {Workman}}, \bibinfo {author} {\bibfnamefont {R.~A.}\ \bibnamefont {Arndt}}, \bibinfo {author} {\bibfnamefont {W.~J.}\ \bibnamefont {Briscoe}}, \bibinfo {author} {\bibfnamefont {M.~W.}\ \bibnamefont {Paris}},\ and\ \bibinfo {author} {\bibfnamefont {I.~I.}\ \bibnamefont {Strakovsky}},\ }\bibfield  {title} {\bibinfo {title} {{Parameterization dependence of T matrix poles and eigenphases from a fit to $\pi$N elastic scattering data}},\ }\href {https://doi.org/10.1103/PhysRevC.86.035202} {\bibfield  {journal} {\bibinfo  {journal} {Phys. Rev. C}\ }\textbf {\bibinfo {volume} {86}},\ \bibinfo {pages} {035202} (\bibinfo {year} {2012})},\ \Eprint {https://arxiv.org/abs/1204.2277} {arXiv:1204.2277 [hep-ph]} \BibitemShut {NoStop}%
\bibitem [{BnG()}]{BnGa-web}%
  \BibitemOpen
  \href@noop {} {\bibinfo {title} {{\text https://pwa.hiskp.uni-bonn.de/}}}\BibitemShut {NoStop}%
\bibitem [{\citenamefont {Ajaka}\ \emph {et~al.}(2007)\citenamefont {Ajaka} \emph {et~al.}}]{Ajaka:2007zz}%
  \BibitemOpen
  \bibfield  {author} {\bibinfo {author} {\bibfnamefont {J.}~\bibnamefont {Ajaka}} \emph {et~al.},\ }\bibfield  {title} {\bibinfo {title} {{Double $\pi^0$ photoproduction on the neutron at GRAAL}},\ }\href {https://doi.org/10.1016/j.physletb.2007.06.009} {\bibfield  {journal} {\bibinfo  {journal} {Phys. Lett. B}\ }\textbf {\bibinfo {volume} {651}},\ \bibinfo {pages} {108} (\bibinfo {year} {2007})}\BibitemShut {NoStop}%
\bibitem [{\citenamefont {Anisovich}\ and\ \citenamefont {Sarantsev}(2006)}]{Anisovich:2006bc}%
  \BibitemOpen
  \bibfield  {author} {\bibinfo {author} {\bibfnamefont {A.~V.}\ \bibnamefont {Anisovich}}\ and\ \bibinfo {author} {\bibfnamefont {A.~V.}\ \bibnamefont {Sarantsev}},\ }\bibfield  {title} {\bibinfo {title} {{Partial decay widths of baryons in the spin-momentum operator expansion method}},\ }\href {https://doi.org/10.1140/epja/i2006-10102-1} {\bibfield  {journal} {\bibinfo  {journal} {Eur. Phys. J. A}\ }\textbf {\bibinfo {volume} {30}},\ \bibinfo {pages} {427} (\bibinfo {year} {2006})},\ \Eprint {https://arxiv.org/abs/hep-ph/0605135} {arXiv:hep-ph/0605135} \BibitemShut {NoStop}%
\bibitem [{\citenamefont {Aitchison}(1972)}]{Aitchison:1972ay}%
  \BibitemOpen
  \bibfield  {author} {\bibinfo {author} {\bibfnamefont {I.~J.~R.}\ \bibnamefont {Aitchison}},\ }\bibfield  {title} {\bibinfo {title} {{K-matrix formalism for overlappimg resonances}},\ }\href {https://doi.org/10.1016/0375-9474(72)90305-3} {\bibfield  {journal} {\bibinfo  {journal} {Nucl. Phys. A}\ }\textbf {\bibinfo {volume} {189}},\ \bibinfo {pages} {417} (\bibinfo {year} {1972})}\BibitemShut {NoStop}%
\bibitem [{\citenamefont {Chung}\ \emph {et~al.}(1995)\citenamefont {Chung} \emph {et~al.}}]{Chung:1995dx}%
  \BibitemOpen
  \bibfield  {author} {\bibinfo {author} {\bibfnamefont {S.~U.}\ \bibnamefont {Chung}} \emph {et~al.},\ }\bibfield  {title} {\bibinfo {title} {{Partial wave analysis in K matrix formalism}},\ }\href {https://doi.org/10.1002/andp.19955070504} {\bibfield  {journal} {\bibinfo  {journal} {Annalen Phys.}\ }\textbf {\bibinfo {volume} {4}},\ \bibinfo {pages} {404} (\bibinfo {year} {1995})}\BibitemShut {NoStop}%
\bibitem [{\citenamefont {Peters}(2006)}]{Peters:2004qw}%
  \BibitemOpen
  \bibfield  {author} {\bibinfo {author} {\bibfnamefont {K.~J.}\ \bibnamefont {Peters}},\ }\bibfield  {title} {\bibinfo {title} {{A Primer on partial wave analysis}},\ }\href {https://doi.org/10.1142/S0217751X06034811} {\bibfield  {journal} {\bibinfo  {journal} {Int. J. Mod. Phys. A}\ }\textbf {\bibinfo {volume} {21}},\ \bibinfo {pages} {5618} (\bibinfo {year} {2006})},\ \Eprint {https://arxiv.org/abs/hep-ph/0412069} {arXiv:hep-ph/0412069} \BibitemShut {NoStop}%
\bibitem [{\citenamefont {Sarantsev}\ \emph {et~al.}(2009)\citenamefont {Sarantsev}, \citenamefont {Anisovich}, \citenamefont {Nikonov},\ and\ \citenamefont {Schmieden}}]{Sarantsev:2008ar}%
  \BibitemOpen
  \bibfield  {author} {\bibinfo {author} {\bibfnamefont {A.~V.}\ \bibnamefont {Sarantsev}}, \bibinfo {author} {\bibfnamefont {A.~V.}\ \bibnamefont {Anisovich}}, \bibinfo {author} {\bibfnamefont {V.~A.}\ \bibnamefont {Nikonov}},\ and\ \bibinfo {author} {\bibfnamefont {H.}~\bibnamefont {Schmieden}},\ }\bibfield  {title} {\bibinfo {title} {{Polarization degrees of freedom in near-threshold photoproduction of omega mesons in the pi0 gamma decay channel}},\ }\href {https://doi.org/10.1140/epja/i2008-10696-0} {\bibfield  {journal} {\bibinfo  {journal} {Eur. Phys. J. A}\ }\textbf {\bibinfo {volume} {39}},\ \bibinfo {pages} {61} (\bibinfo {year} {2009})},\ \Eprint {https://arxiv.org/abs/0806.4477} {arXiv:0806.4477 [hep-ex]} \BibitemShut {NoStop}%
\bibitem [{\citenamefont {Anisovich}\ \emph {et~al.}(2005)\citenamefont {Anisovich}, \citenamefont {Klempt}, \citenamefont {Sarantsev},\ and\ \citenamefont {Thoma}}]{Anisovich:2004zz}%
  \BibitemOpen
  \bibfield  {author} {\bibinfo {author} {\bibfnamefont {A.}~\bibnamefont {Anisovich}}, \bibinfo {author} {\bibfnamefont {E.}~\bibnamefont {Klempt}}, \bibinfo {author} {\bibfnamefont {A.}~\bibnamefont {Sarantsev}},\ and\ \bibinfo {author} {\bibfnamefont {U.}~\bibnamefont {Thoma}},\ }\bibfield  {title} {\bibinfo {title} {{Partial wave decomposition of pion and photoproduction amplitudes}},\ }\href {https://doi.org/10.1140/epja/i2004-10125-6} {\bibfield  {journal} {\bibinfo  {journal} {Eur. Phys. J. A}\ }\textbf {\bibinfo {volume} {24}},\ \bibinfo {pages} {111} (\bibinfo {year} {2005})},\ \Eprint {https://arxiv.org/abs/hep-ph/0407211} {arXiv:hep-ph/0407211} \BibitemShut {NoStop}%
\bibitem [{\citenamefont {Burkert}\ \emph {et~al.}(2023)\citenamefont {Burkert} \emph {et~al.}}]{Burkert:2022bqo}%
  \BibitemOpen
  \bibfield  {author} {\bibinfo {author} {\bibfnamefont {V.}~\bibnamefont {Burkert}} \emph {et~al.},\ }\bibfield  {title} {\bibinfo {title} {{Note on the definitions of branching ratios of overlapping resonances}},\ }\href {https://doi.org/10.1016/j.physletb.2023.138070} {\bibfield  {journal} {\bibinfo  {journal} {Phys. Lett. B}\ }\textbf {\bibinfo {volume} {844}},\ \bibinfo {pages} {138070} (\bibinfo {year} {2023})},\ \Eprint {https://arxiv.org/abs/2207.08472} {arXiv:2207.08472 [hep-ph]} \BibitemShut {NoStop}%
\bibitem [{\citenamefont {Von~Hippel}\ and\ \citenamefont {Quigg}(1972)}]{VonHippel:1972fg}%
  \BibitemOpen
  \bibfield  {author} {\bibinfo {author} {\bibfnamefont {F.}~\bibnamefont {Von~Hippel}}\ and\ \bibinfo {author} {\bibfnamefont {C.}~\bibnamefont {Quigg}},\ }\bibfield  {title} {\bibinfo {title} {{Centrifugal-barrier effects in resonance partial decay widths, shapes, and production amplitudes}},\ }\href {https://doi.org/10.1103/PhysRevD.5.624} {\bibfield  {journal} {\bibinfo  {journal} {Phys. Rev. D}\ }\textbf {\bibinfo {volume} {5}},\ \bibinfo {pages} {624} (\bibinfo {year} {1972})}\BibitemShut {NoStop}%
\bibitem [{\citenamefont {Heuser}\ \emph {et~al.}(2024)\citenamefont {Heuser} \emph {et~al.}}]{Heuser:2024biq}%
  \BibitemOpen
  \bibfield  {author} {\bibinfo {author} {\bibfnamefont {L.~A.}\ \bibnamefont {Heuser}} \emph {et~al.},\ }\bibfield  {title} {\bibinfo {title} {{From pole parameters to line shapes and branching ratios}},\ }\href {https://doi.org/10.1140/epjc/s10052-024-12884-6} {\bibfield  {journal} {\bibinfo  {journal} {Eur. Phys. J. C}\ }\textbf {\bibinfo {volume} {84}},\ \bibinfo {pages} {599} (\bibinfo {year} {2024})},\ \Eprint {https://arxiv.org/abs/2403.15539} {arXiv:2403.15539 [hep-ph]} \BibitemShut {NoStop}%
\bibitem [{\citenamefont {Erbe}\ \emph {et~al.}(1968{\natexlab{b}})\citenamefont {Erbe} \emph {et~al.}}]{Aachen-Berlin-Bonn-Hamburg-Heidelberg-Munich:1968rzt}%
  \BibitemOpen
  \bibfield  {author} {\bibinfo {author} {\bibfnamefont {R.}~\bibnamefont {Erbe}} \emph {et~al.} (\bibinfo {collaboration} {Aachen-Berlin-Bonn-Hamburg-Heidelberg-Munich}),\ }\bibfield  {title} {\bibinfo {title} {{Photoproduction of Meson and Baryon Resonances at Energies up to 5.8 GEV}},\ }\href {https://doi.org/10.1103/PhysRev.175.1669} {\bibfield  {journal} {\bibinfo  {journal} {Phys. Rev.}\ }\textbf {\bibinfo {volume} {175}},\ \bibinfo {pages} {1669} (\bibinfo {year} {1968}{\natexlab{b}})}\BibitemShut {NoStop}%
\bibitem [{SAI()}]{SAID}%
  \BibitemOpen
  \href@noop {} {\bibinfo {title} {https://gwdac.phys.gwu.edu/}}\BibitemShut {NoStop}%
\bibitem [{\citenamefont {Roy}\ \emph {et~al.}(2025)\citenamefont {Roy} \emph {et~al.}}]{CLAS:2025ogo}%
  \BibitemOpen
  \bibfield  {author} {\bibinfo {author} {\bibfnamefont {P.}~\bibnamefont {Roy}} \emph {et~al.} (\bibinfo {collaboration} {CLAS}),\ }\bibfield  {title} {\bibinfo {title} {{Measurement of single- and double-polarization observables in the photoproduction of $\pi^+\pi^-${\textasciitilde}meson pairs off the proton using CLAS at Jefferson Laboratory}},\ }\href@noop {} {\  (\bibinfo {year} {2025})},\ \Eprint {https://arxiv.org/abs/2504.21119} {arXiv:2504.21119 [nucl-ex]} \BibitemShut {NoStop}%
\bibitem [{\citenamefont {Mokeev}\ \emph {et~al.}(2023)\citenamefont {Mokeev} \emph {et~al.}}]{Mokeev:2023zhq}%
  \BibitemOpen
  \bibfield  {author} {\bibinfo {author} {\bibfnamefont {V.~I.}\ \bibnamefont {Mokeev}} \emph {et~al.},\ }\bibfield  {title} {\bibinfo {title} {{First Results on Nucleon Resonance Electroexcitation Amplitudes from $ep\to e'\pi^+\pi^-p'$ Cross Sections at $W$ = 1.4-1.7 GeV and $Q^2$ = 2.0-5.0 GeV$^2$}},\ }\href {https://doi.org/10.1103/PhysRevC.108.025204} {\bibfield  {journal} {\bibinfo  {journal} {Phys. Rev. C}\ }\textbf {\bibinfo {volume} {108}},\ \bibinfo {pages} {025204} (\bibinfo {year} {2023})},\ \Eprint {https://arxiv.org/abs/2306.13777} {arXiv:2306.13777 [nucl-ex]} \BibitemShut {NoStop}%
\bibitem [{\citenamefont {Eichmann}\ and\ \citenamefont {Ramalho}(2018)}]{Eichmann:2018ytt}%
  \BibitemOpen
  \bibfield  {author} {\bibinfo {author} {\bibfnamefont {G.}~\bibnamefont {Eichmann}}\ and\ \bibinfo {author} {\bibfnamefont {G.}~\bibnamefont {Ramalho}},\ }\bibfield  {title} {\bibinfo {title} {{Nucleon resonances in Compton scattering}},\ }\href {https://doi.org/10.1103/PhysRevD.98.093007} {\bibfield  {journal} {\bibinfo  {journal} {Phys. Rev. D}\ }\textbf {\bibinfo {volume} {98}},\ \bibinfo {pages} {093007} (\bibinfo {year} {2018})},\ \Eprint {https://arxiv.org/abs/1806.04579} {arXiv:1806.04579 [hep-ph]} \BibitemShut {NoStop}%
\bibitem [{\citenamefont {Hoferichter}\ \emph {et~al.}(2024)\citenamefont {Hoferichter}, \citenamefont {de~Elvira}, \citenamefont {Kubis},\ and\ \citenamefont {Mei\ss{}ner}}]{Hoferichter:2023mgy}%
  \BibitemOpen
  \bibfield  {author} {\bibinfo {author} {\bibfnamefont {M.}~\bibnamefont {Hoferichter}}, \bibinfo {author} {\bibfnamefont {J.~R.}\ \bibnamefont {de~Elvira}}, \bibinfo {author} {\bibfnamefont {B.}~\bibnamefont {Kubis}},\ and\ \bibinfo {author} {\bibfnamefont {U.-G.}\ \bibnamefont {Mei\ss{}ner}},\ }\bibfield  {title} {\bibinfo {title} {{Nucleon resonance parameters from Roy\textendash{}Steiner equations}},\ }\href {https://doi.org/10.1016/j.physletb.2024.138698} {\bibfield  {journal} {\bibinfo  {journal} {Phys. Lett. B}\ }\textbf {\bibinfo {volume} {853}},\ \bibinfo {pages} {138698} (\bibinfo {year} {2024})},\ \Eprint {https://arxiv.org/abs/2312.15015} {arXiv:2312.15015 [hep-ph]} \BibitemShut {NoStop}%
\bibitem [{\citenamefont {Höhler}(1993)}]{Hohler:1993lbk}%
  \BibitemOpen
  \bibfield  {author} {\bibinfo {author} {\bibfnamefont {G.}~\bibnamefont {Höhler}},\ }\bibfield  {title} {\bibinfo {title} {{Determination of $\pi N$ resonance pole parameters}},\ }\href@noop {} {\bibfield  {journal} {\bibinfo  {journal} {PiN Newslett.}\ }\textbf {\bibinfo {volume} {1993}},\ \bibinfo {pages} {1} (\bibinfo {year} {1993})}\BibitemShut {NoStop}%
\bibitem [{\citenamefont {Cutkosky}\ \emph {et~al.}(1980)\citenamefont {Cutkosky} \emph {et~al.}}]{Cutkosky:1980rh}%
  \BibitemOpen
  \bibfield  {author} {\bibinfo {author} {\bibfnamefont {R.~E.}\ \bibnamefont {Cutkosky}} \emph {et~al.},\ }\bibfield  {title} {\bibinfo {title} {{Pion - Nucleon Partial Wave Analysis.\hfill }},\ }in\ \href@noop {} {\emph {\bibinfo {booktitle} {{4th International Conference on Baryon Resonances}}}}\ (\bibinfo {year} {1980})\ p.~\bibinfo {pages} {19},\ \bibinfo {note} {14-16 July 1980. Toronto, Ontario, Canada}\BibitemShut {NoStop}%
\end{thebibliography}%

\end{document}